\documentclass[trackchanges, twocolumn]{aastex701}
\usepackage{subcaption}
\usepackage{booktabs}
\usepackage{multirow}
\usepackage{tabularx}
\usepackage{array}
\usepackage{amsmath}
\usepackage{placeins}

\newcommand{\nii}{[{\sc N\,ii}]}
\newcommand{\sii}{[{\sc S\,ii}]}

\begin{document}

\title{High Resolution Spectro-Polarimetric Observations of the Optical Filaments surrounding NGC 1275 with ESPaDOnS}

\author[0000-0002-2478-5119]{Benjamin Vigneron}
\affiliation{Département de Physique, Université de Montréal, Succ. Centre-Ville, Montréal, Québec, H3C 3J7, Canada}
\affiliation{Centre de recherche en astrophysique du Québec (Astroquébec)}
\email{benjamin.vigneron@umontreal.ca}

\author[0000-0001-7271-7340]{Julie Hlavacek-Larrondo}
\affiliation{Département de Physique, Université de Montréal, Succ. Centre-Ville, Montréal, Québec, H3C 3J7, Canada}
\affiliation{Centre de recherche en astrophysique du Québec (Astroquébec)}
\email{j.larrondo@umontreal.ca}

\author[0009-0002-9966-6517]{Luca Fabiani}
\affiliation{Département de Physique, Université de Montréal, Succ. Centre-Ville, Montréal, Québec, H3C 3J7, Canada}
\email{luca.fabiani@umontreal.ca}

\author[0000-0002-8786-8499]{Lison Malo}
\affiliation{Département de Physique, Université de Montréal, Succ. Centre-Ville, Montréal, Québec, H3C 3J7, Canada}
\email{lison.malo@umontreal.ca}

\author[0000-0003-2001-1076]{Carter Lee Rhea}
\email{carter.rhea@dragonfly1000.com}
\affiliation{Dragonfly Focused Research Organization, 150 Washington Avenue, Santa Fe, NM 87501, USA}

\author[0000-0002-2808-0853]{Megan Donahue}
\affiliation{Department of Physics and Astronomy, Michigan State University, East Lansing, MI 48824, USA}
\email{donahu42@msu.edu}

\author[0000-0003-2405-7258]{Jorge Karolt Barrera Ballesteros}
\affiliation{Instituto de Astronomía, Universidad Nacional Autónoma de México, A.P. 70-264, 04510 CDMX, México}
\email{jkbarrerab@astro.unam.mx}

\author[0000-0002-3173-1098]{Hyunseop Choi}
\affiliation{Department of Astronomy, University of Michigan, 1085 S. University Ave., Ann Arbor, MI 48109, USA}
\email{joechoi@umich.edu}

\author[0000-0001-7548-0190]{Nadine Manset}
\affiliation{Canada-France-Hawaii Telescope, 65-1238 Mamalahoa Hwy, Kamuela, Hawaii 96743, USA}
\email{manset@cfht.hawaii.edu}





\begin{abstract}

We analyze archival high spectral resolution ESPaDOnS spectropolarimetric observations (R = $\lambda/\Delta\lambda$ = 68000) of three targeted locations within the filamentary nebula surrounding NGC 1275. We present here the analysis of the main filaments emission lines, namely: \nii$\lambda6584$, H$\alpha$, and \nii$\lambda6548$, across our three pointings displaying clear detectable emission, with one also showing additional \sii$\lambda6716$ and \sii$\lambda6731$ emission line doublet. Using a Markov Chain Monte-Carlo (MCMC) algorithm and thanks to the exceptionally high spectral resolution of the observations, we resolve for the first time each of the emission lines within one of the pointings into a set of three emission components with a remarkable velocity dispersion contrast ($\sigma_1 \sim 25$ km/s, $\sigma_2 \sim 90$ km/s and $\sigma_3 \sim 208$ km/s). We also perform flux ratios of the three detected emission components, which reveal differences in electron density and ionization sources. Consequently, the ESPaDOnS observations appear to reveal for the first time the overlap of an outer quiescent filament and a central turbulent filament likely affected by the AGN activity at a distance of $\sim 6.8''$ ($\sim 2.5$ kpc) from the central BCG. Finally, ESPaDOnS polarimetric data of one of the pointing allows us to confirm an average linear polarization degree of $\sim 0.15-0.14\%$ and an angle of polarization of $5.3\pm25.7^\circ$ for the optical emission of the filaments, as well as polarization dips associated with the H$\alpha$ emission line and its multiple emission components. We interpret these features and measurements as tentative polarization due to dust scattering.

\end{abstract}

\keywords{Galaxies: NGC 1275 - Galaxies: clusters: individual: Perseus cluster}

\clearpage
\section{Introduction} \label{sec:intro}




Galaxy clusters are large gravitationally bound systems that host hundreds to thousands of galaxies (e.g. \citealt{bahcall_clusters_1977}, \citealt{abell_catalog_1989}). However, galaxies account for only a small fraction ($\sim 5\%$) of the cluster's total mass, while dark matter and ionised gas dominate the overall mass distribution (e.g. \citealt{sand_dark_2004}, \citealt{voigt_galaxy_2006}). This ionised gas, called the intra-cluster medium (ICM), is a plasma at extremely high temperature ($\sim 10^7-10^8$ K) but relatively low overall density  ($\sim 10^{-1}-10^{-3} \text{ cm}^{-3}$ in the core and with a mean cluster gas density of $1.5\times10^{-4}\text{ cm}^{-3}$ - e.g. \citealt{cavaliere_extragalactic_1971}, \citealt{gursky_detection_1971}). The ICM emits in X-rays largely due to the bremsstrahlung radiation produced by interactions between ions and free electrons that compose this environment, resulting in Fe forbidden lines as well as many lower-energy lines including  C, N, O, Ne, Mg, Si, S, Ar, and Ca (e.g. \citealt{sarazin1986x}).

Galaxy clusters typically host a dominant central galaxy, known as the brightest cluster galaxy (BCG), often located near the cluster's center. These BCGs are oftentimes associated with extended optical nebula consisting of ionised and molecular gas filaments (e.g. \citealt{crawford_rosat_1999}, \citealt{mcdonald_origin_2010}, \citealt{salome_very_2011}). These filaments, tracing a complex multiphase environment emitting in optical, X-ray and radio, probe several orders of magnitude in gas temperature ($10-10^7$ K - \citealt{salome_very_2011}, \citealt{olivares2025halpha}). 
Optically emitting filaments of ionised gas trace temperatures of $\sim 10^4$ K and display generally high H$\alpha$ luminosities up to $10^{42}\text{ erg.s}^{-1}$ (e.g. \citealt{conselice2001nature})

In this paper, we focus on the optically-emitting filamentary nebula of ionised gas surrounding NGC 1275 ($z = 0.017284$ - \citealt{hitomi_collaboration_atmospheric_2018}), BCG of the close by Perseus cluster, which has been observed exhaustively at various wavelengths (see \citealt{forman_observations_1972}, \citealt{fabian_wide_2011}, \citealt{salome_very_2011}, \citealt{lim_molecular_2012}, \citealt{nagai_alma_2019}). Additionally, NGC 1275 displays one of the most extended filamentary nebula ever observed with a size of roughly $80$ kpc $ \times \text{ } 50$ kpc (e.g. \citealt{mcnamara_optical_1996}, \citealt{conselice2001nature}, \citealt{hamer_optical_2016}). NGC 1275 was first observed by \cite{minkowski_optical_1957}, \cite{lynds1970improved}, and \cite{rubin1977new} revealing the presence of two emission structures separated by their respective velocities. A redshifted high-velocity feature ($\sim$ 8200 km/s) was associated to a foreground, dusty, spiral galaxy trapped in the cluster potential and falling onto the BCG (e.g. \citealt{yu2015high, rhea2025multiwavelength}), while a low-velocity structure ($\sim$ 5200 km/s) corresponded to the vast filamentary nebula of ionised gas surrounding NGC 1275. Afterwards, \textit{Hubble Space Telescope} (HST) observations of this low-velocity structure revealed the thread-like appearance of these filaments (e.g. \citealt{fabian_magnetic_2008}), while soft X-ray emission appeared associated with specific filaments using \textit{Chandra X-Ray Observatory} observations (e.g. \citealt{fabian_relationship_2003})



Nevertheless, the first detailed observations of the nebula were performed by \cite{conselice2001nature} with high-resolution imaging, integral field, and long-slit spectroscopy using the National Optical Astronomy Observatory in consortium with the University of Wisconsin-Madison, Indiana University and Yale University as well as the Kitt Peak National Observatory (WIYN \& KPNO). These observations brought forth the first velocity map of the central $\sim 45 \text{ arcsec}$ ($\sim 16.5$ kpc) of the nebula. Additional observations from the Gemini Multi-Object Spectrograph along six slits carefully positioned along certain filaments showed evidence of outflowing gas and flow patterns (e.g. \citealt{hatch_origin_2006}). In 2018 and again in 2020, the filamentary nebula was then imaged with SITELLE (\textit{Spectromètre Imageur à Transformée de Fourier pour l’Etude en Long et en Large de raies d’Emission}), an imaging Fourier transform spectrometer (e.g. \citealt{2019MNRAS.485.3930D}) installed at the Canada-France-Hawaï Telescope (CFHT) possessing an extremely large field of view ($11' \times 11'$) capable of observing the nebula in its entirety and extracting the emission spectra of the filaments on a pixel scale of $0.32''$. These observations at low and high spectral resolution respectively (R = 1800 and 7000, corresponding to an instrumental velocity resolution of 166.6 km/s and 42.8 km/s respectively, with a theoretical lower bound on the velocity uncertainty of $\sim 0.2$ km/s) revealed for the first time the complete flux, bulk velocity and velocity dispersion maps of the filaments (see \citealt{gendron-marsolais_revealing_2018, vigneron2024high}). These observations also demonstrated that, despite relatively high spectral resolution, multiple emission components could not be properly disentangled within the emission spectra of the filaments.

Additionally, these optical filaments of ionised gas display a relatively extended lifetime and do not appear to significantly collapse, as limited star formation is observed within them (e.g. \citealt{canning2010star}, \citealt{canning_filamentary_2014}). A proposed model to explain their relatively long lifetime and stability in this complex environment would be through the presence of magnetic fields induced by magnetic pressure, which would allow the stabilization of the filaments and therefore prevent their collapse (e.g. \citealt{fabian_magnetic_2008}). Polarimetric observations of the optical filaments surrounding the BCGs M87, NGC 4696, Hydra A and PKS0745-19 with slit spectro-polarimetry using the \textit{Very Large Telescope} (VLT) UT1 FORS2 instrument, appear to indicate that only a low linear polarization degree ($<0.1\%$) could be observed within the filaments, suggesting that such magnetic fields could be present but undetected (e.g. \citealt{sparks2013polarization}). Recent optical spectropolarimetric observations of NGC 1275's AGN using a 1.5''-wide and 15''-long slit indicate that a higher but variable linear polarization degree is detectable ($\sim 1-2\%$) and appears associated with the AGN's jets, as the measured polarization angle and physical angle of the jets seem to correlate (e.g. \citealt{marin2025spectropolarimetry}).



In this paper, we analyze extremely high spectral resolution (R = 68000) spectropolarimetric observations of the filamentary nebula surrounding NGC 1275 obtained with the spectropolarimeter ESPaDOnS installed on the CFHT at the Mauna Kea Observatory in Hawaï. 
This allows us to probe the main emission lines of three pointings within the filaments, located at a distance of $\sim 6 - 8''$ from the central galaxy, down to an instrumental velocity resolution of $\sim 4.4$ km/s. Additionally, these ESPaDOnS observations explore the polarimetric profile of the filaments surrounding NGC 1275. Such analysis has only been carried out around other BCGs' filaments and NGC 1275's AGN (see \citealt{sparks2013polarization}, \citealt{marin2025spectropolarimetry}) but will offer us a first glimpse of the possible polarization present within the optical spectral emission of the extended filaments surrounding NGC 1275.

In Section 2, we present the spectropolarimetric ESPaDOnS observations and the analysis of the main emission lines observed in the spectral data, followed by a study of the polarimetric features associated with these emission lines. In Section 3, we present and discuss our results regarding the multi-component fitting of the emission lines as well as the tentative features detected within the polarimetric data associated with the main emission lines.

To directly compare our results to those of previous SITELLE observations \cite{gendron-marsolais_revealing_2018}, \cite{vigneron2024high} and the Hitomi \cite{hitomi_collaboration_atmospheric_2018}, we also adopt for NGC 1275 a redshift of $z=0.017284$ for the stellar component, which implies an angular scale of $21.2 \; \text{kpc arcmin}^{-1}$. This $z$ also corresponds to a luminosity distance of $75.5$ Mpc, assuming $H_0 = 69.6 \; \text{km s}^{-1} \text{Mpc}^{-1}$, $\Omega_M = 0.286$ and $\Omega_{vac} = 0.714$. All the errors displayed in our results correspond to the $3\sigma$ confidence level unless otherwise noted. 

\section{Data Analysis} \label{sec:data}

\subsection{Observations with ESPaDOnS \label{sec:obs}}

\begin{figure*}
    \centering
    \includegraphics[width=0.9\linewidth]{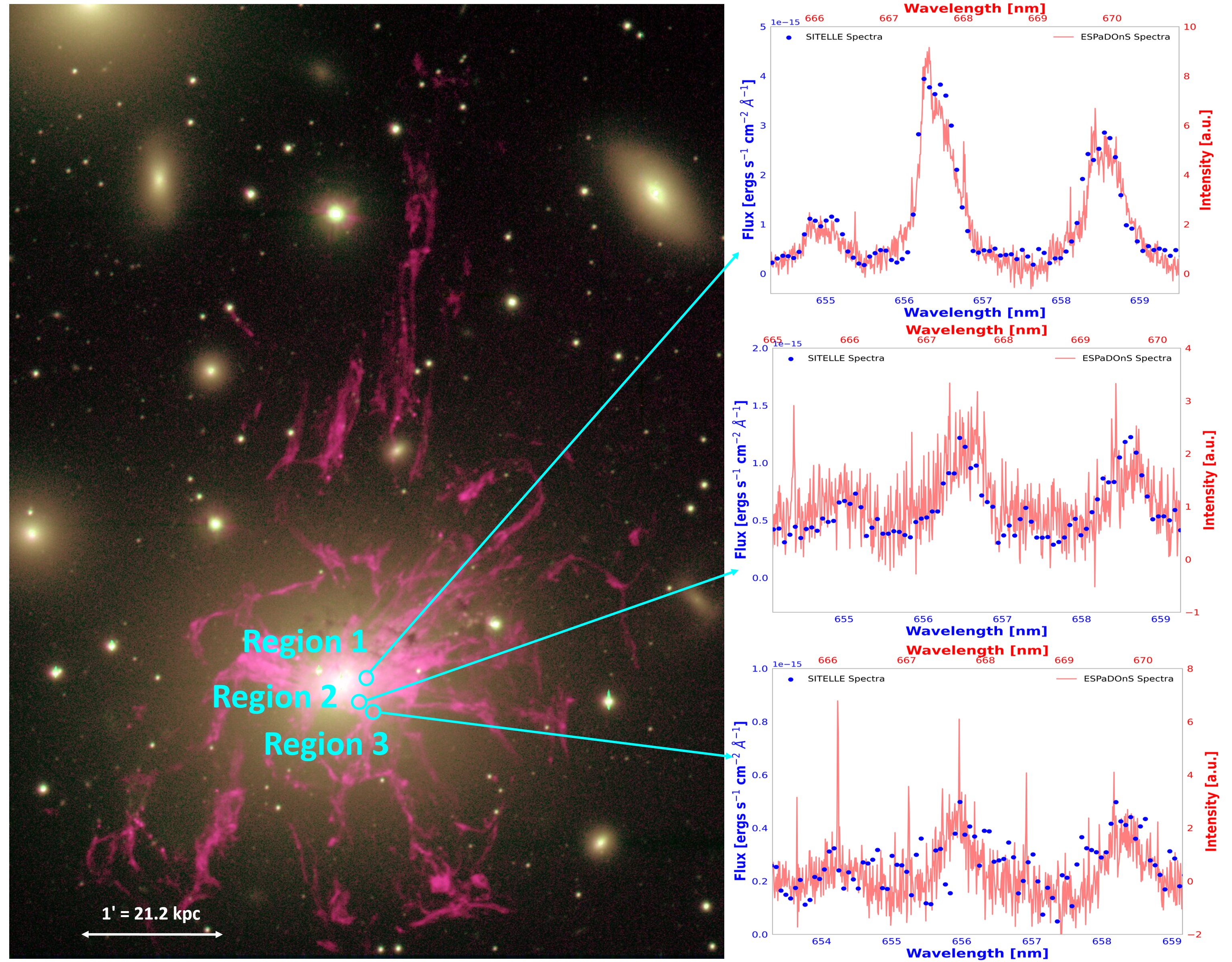}
    \caption{Location of the three ESPaDOnS pointings displaying emission signal observing key locations within the filamentary nebula surrounding NGC 1275 overplotted on WIYS observations (\citealt{conselice2001nature}). The associated ESPaDOnS and SITELLE spectra at high spectral resolution are shown in red and blue respectively with their corresponding location indicated with cyan circles denoted as Region 1, located at RA: 3:19:47.60 and DEC: 41:30:43.2; Region 2, located at RA: 3:19:47.70 and DEC: 41:30:37.2, and Region 3, located at RA: 3:19:47.40, DEC: 41:30:33.9. Since the SITELLE observations are redshift-corrected and the ESPaDOnS observations are not, the corresponding wavelengths axis associated with each dataset are also colour-coded (respectively in blue for SITELLE and red for ESPaDOnS).}
    \label{fig:espadons_pointings}
\end{figure*}

ESPaDOnS is a bench-mounted echelle spectropolarimeter capable of reaching extremely high spectral resolution (up to $R = \frac{\lambda}{\Delta\lambda} \sim 81000 \text{ or } 3.7$ km/s in single target mode and $R = \frac{\lambda}{\Delta\lambda} \sim 68000 \text{ or } 4.4$ km/s in target + sky mode - e.g. \citealt{donati2003espadons}, \citealt{moutou2015news}). Installed on the CFHT telescope, ESPaDOnS possesses a large spectral bandwidth covering the entirety of the optical domain and parts of the infrared and ultraviolet ($370$ nm to $1000$ nm). Despite it relatively small aperture of $\sim 1.6''$ radius, ESPaDOnS is a prime instrument for the spectral study of stars and their polarization properties.

ESPaDOnS has also been used to study more extended targets such as the filamentary nebula surrounding NGC 1275. Observations were taken in November 2010 during Cycle 10B by PIs W. B. Sparks and N. Manset and the data were made public in 2012. The observations targeted five pointings within the filamentary nebula, three of which are located relatively close to the central BCG, one close to the Horseshoe filament and one within the eastern central region of the gaseous structure (see Regions 1, 2 and 3 in Figure \ref{fig:espadons_pointings}, as well as Region 4 and 5 in Appendix \ref{annex:other_pointings}). However, due to the faintness of the outer filaments and the difficulties of offset pointings onto the target, only the three pointings close to the central BCG (Regions 1, 2 and 3) offer a detection of emission lines associated with the filaments. Most notably, one of the pointings displays a significantly higher signal-to-noise ratio (SNR), as well as tentative polarimetric features associated with the emission lines. This pointing, denoted as Region 1 in Figure \ref{fig:espadons_pointings}, will be the main focus of this paper, but dedicated analyses will also be performed on the other two pointings (namely Regions 2 and 3) showing relatively low SNR. We thus display in Figure \ref{fig:espadons_pointings} the location of the three pointings targeted in these observations and analyzed in this paper with their corresponding emission spectra centered on the main emission lines of the filaments, namely \nii$\lambda6584$, H$\alpha$, and \nii$\lambda6548$. We also show in Appendix \ref{annex:other_pointings} the other two pointings and their corresponding spectra displaying no emission signal associated with the surrounding optical filaments.

Five emission lines are detected with a high SNR within the ESPaDOnS spectroscopic observations of the filamentary nebula surrounding NGC 1275, namely in Region 1 located at RA: 3:19:47.60 and DEC: 41:30:43.2: \sii$\lambda6716$, \sii$\lambda6731$, \nii$\lambda6584$, H$\alpha$, and \nii$\lambda6548$. For Regions 2 and 3, located respectively at RA: 3:19:47.70, DEC: 41:30:37.2 and at RA: 3:19:47.40, DEC: 41:30:33.9, only \nii$\lambda6584$, H$\alpha$, and \nii$\lambda6548$ are detected. Moreover, the presence of cosmic rays are also reported within one of the polarization spectra obtained from Region 1. Indeed, through investigation of the ESPaDOnS object files for the measurement of the U Stokes parameters, the presence of cosmic rays were confirmed at wavelengths of $\sim 6672.4 \text{\AA }\text{ } \& \sim 6679.0\text{\AA}$ impacting the H$\alpha$ emission line in the polarization files. These cosmic rays will be indicated throughout the paper in the associated polarization Figures as black dashed lines.

Additionally, a dedicated reduction process is applied on the observational data directly at CFHT using the \texttt{Upena} software, which we will describe briefly here. 

\texttt{Upena} consists of two main components: a quick-analysis process and a final reduction process. The quick-analysis process operates in real time during observations, using pre-sunset calibrations (biases, flat fields, comparison, and Fabry-Perot exposures) to establish geometric and wavelength calibrations. Scientific exposures are reduced as they are captured. Throughout the night, automated checks ensure optimal performance, such as verifying CCD readout noise levels, proper light detection on the CCD, exposure saturation, SNRs, and correct positioning of spectrograph and polarimeter components. The final reduction step reprocesses all data using additional morning flat fields when available and performs polarimetric calculations when applicable. Together, these two components ensure accurate data calibration and analysis throughout the observation process.

In order to compare these observations with similar ones carried out with the SITELLE instrument (R $\sim7000$; see \citealt{vigneron2024high}), we also display the spectra of ESPaDOnS' targeted locations with the corresponding SITELLE spectra of the same regions, where we specifically indicate the pointings analyzed in this paper (see Figure \ref{fig:espadons_pointings} and Appendix \ref{annex:other_pointings}). To do this, we extracted one SITELLE emission spectrum at high spectral resolution for each region corresponding to the ESPaDOnS pointings and aperture thanks to the function \texttt{extract\_spectrum\_region} of the \texttt{LUCI} fitting package for SITELLE observations (\citealt{rhea_luci_2021}).


\subsection{Emission Line Fitting} \label{sec:fitting}

\begin{figure*}[hbt!]
\begin{subfigure}{.495\linewidth}
  \includegraphics[width=\linewidth]{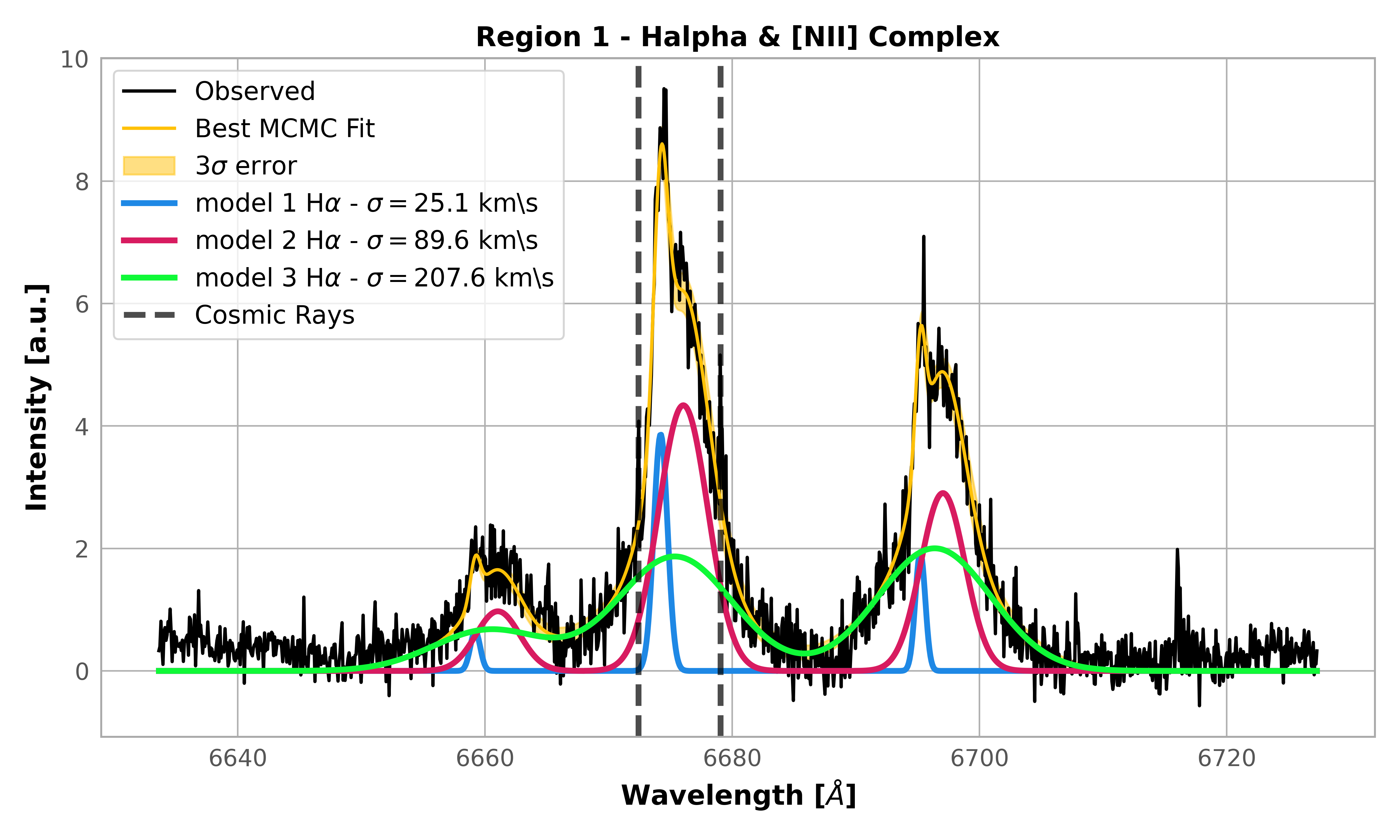}
  \label{fit_halpha}
\end{subfigure}\hfill 
~ 
\begin{subfigure}{.495\linewidth}
  \includegraphics[width=\linewidth]{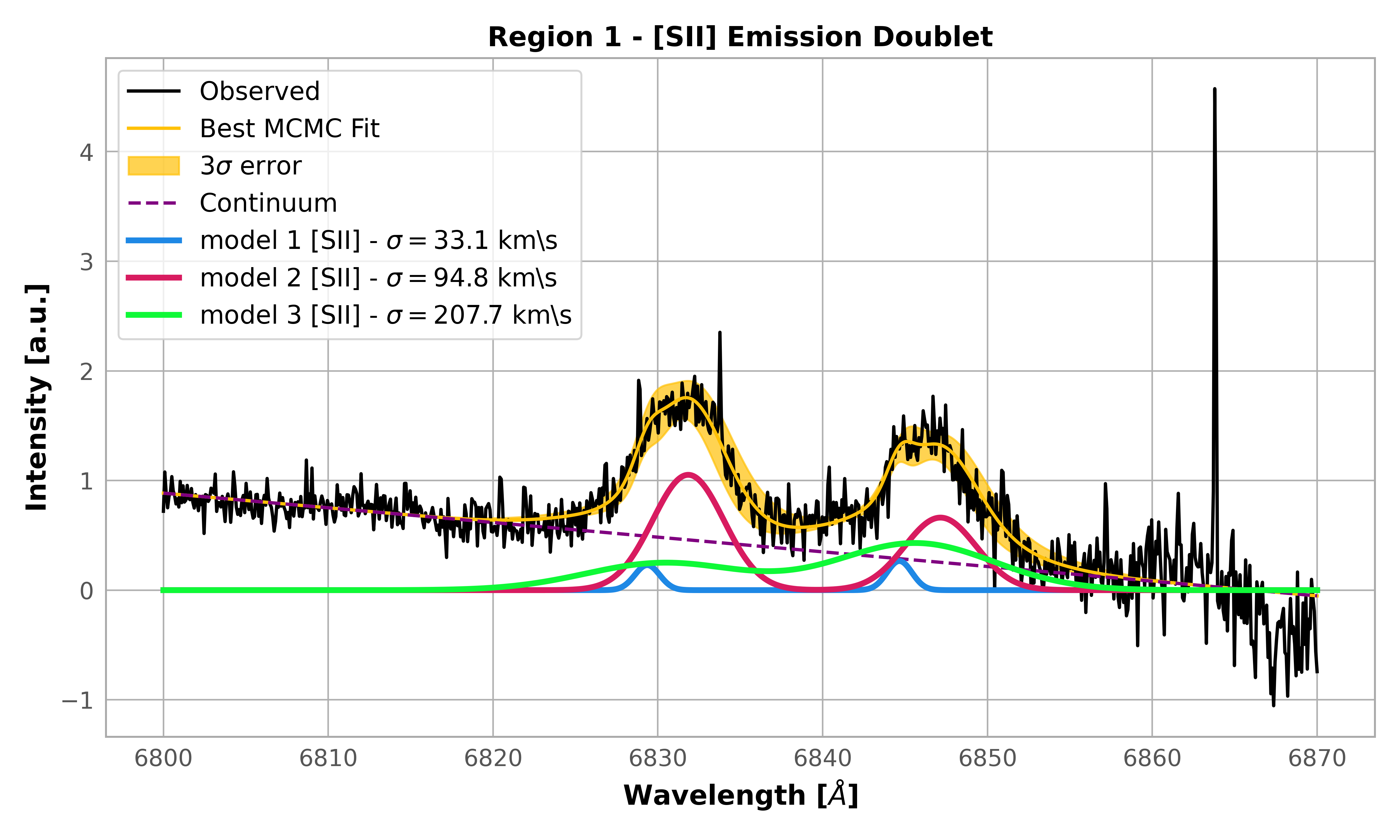}
  \label{fit_sii}
\end{subfigure}

\medskip 
\begin{subfigure}{.495\linewidth}
  \includegraphics[width=\linewidth]{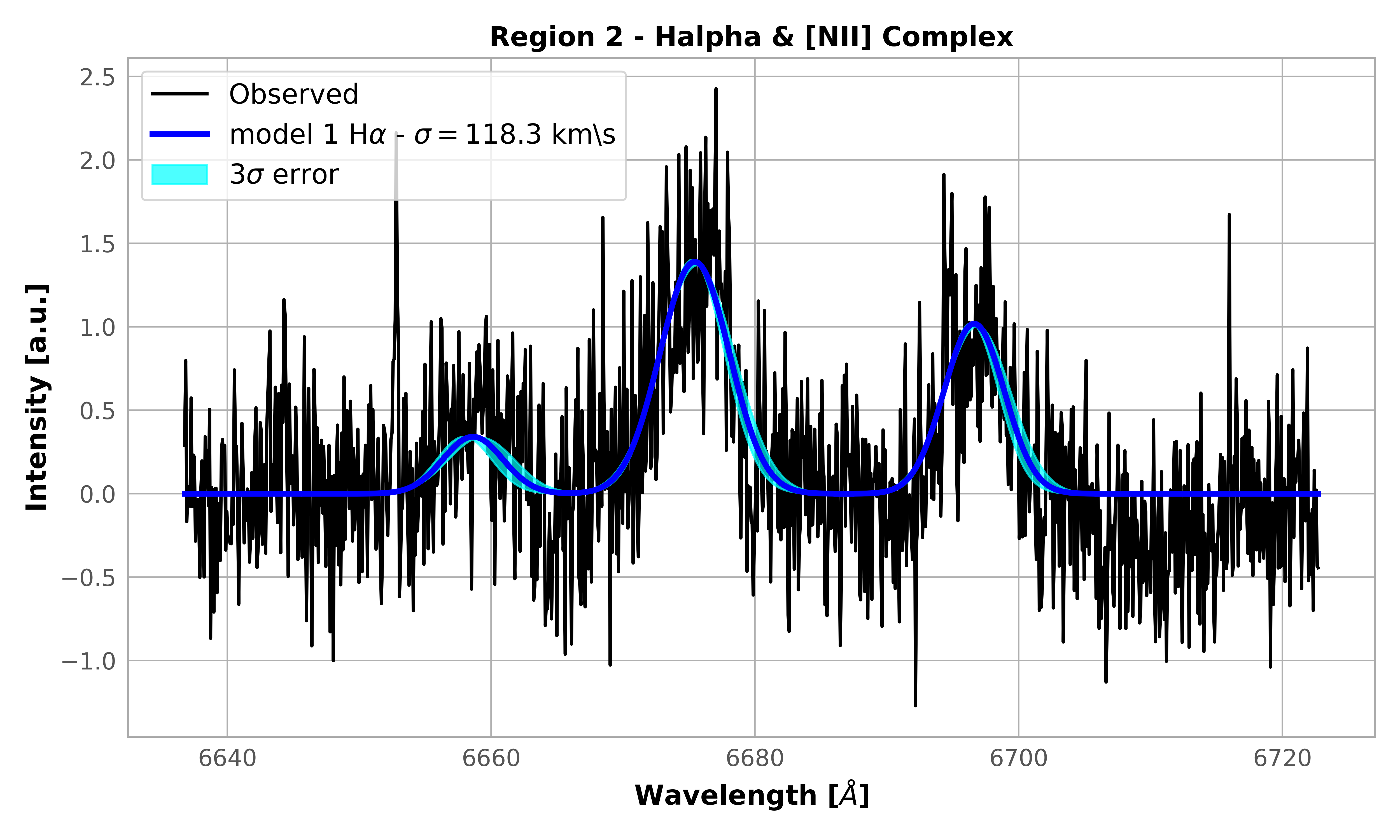}
  \label{pointing_1_fit}
\end{subfigure}\hfill 
\begin{subfigure}{.495\linewidth}
  \includegraphics[width=\linewidth]{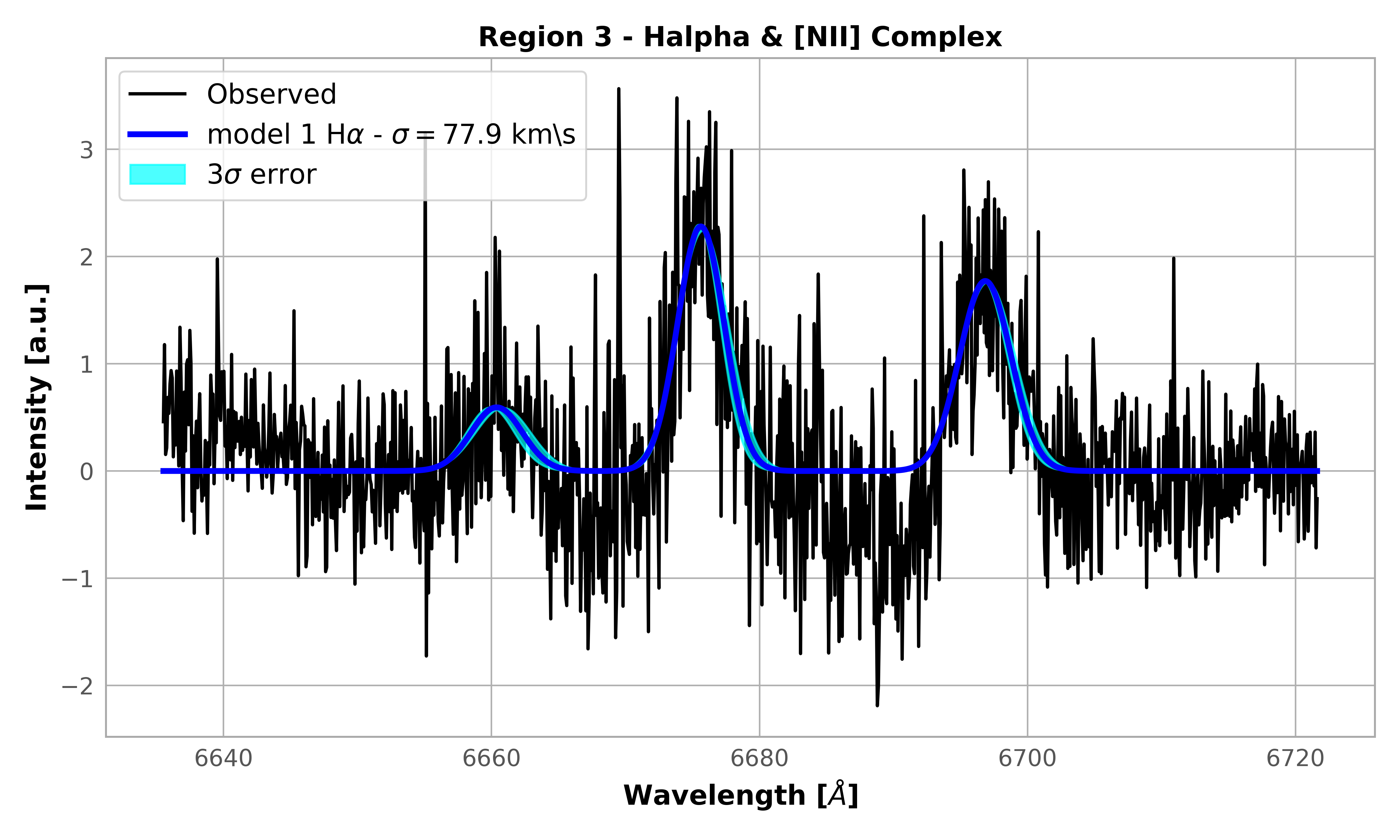}
  \label{pointing_2_fit}
\end{subfigure}

\caption{Top Left: MCMC best fitting result for the H$\alpha$, \nii$\lambda6548$ and \nii$\lambda6584$ emission lines of the first pointing indicating three emission components: a narrow emission with a velocity dispersion of $\sim 25.1$ km/s, an intermediate component with a velocity dispersion of $\sim 89.6$ km/s and a wide component with a velocity dispersion of $\sim 207.6$ km/s for the H$\alpha$ emission line specifically. Top Right: MCMC best fitting result for the \sii$\lambda6716$ and \sii$\lambda6731$ emission lines of the first pointing indicating three emission components: a narrow emission with a velocity dispersion of $\sim 33.1$ km/s, an intermediate component with a velocity dispersion of $\sim 94.8$ km/s and a wide component with a velocity dispersion of $\sim 207.7$ km/s. Bottom Left: MCMC best fitting result for the H$\alpha$, \nii$\lambda6548$ and \nii$\lambda6584$ emission lines of the second pointing indicating a single emission component with a velocity dispersion of $\sim 118.3$ km/s for the H$\alpha$ emission line specifically. Bottom Right: MCMC best fitting result for the H$\alpha$, \nii$\lambda6548$ and \nii$\lambda6584$ emission lines of the third pointing indicating a single emission component with a velocity dispersion of $\sim 78.0$ km/s for the H$\alpha$ emission line specifically.}
\label{fig:fitted_spectra}
\end{figure*}

\begin{deluxetable*}{l l l r r r}
\tablecaption{Emission-Line Best-Fit Parameters and $3\sigma$ Errors from MCMC Modeling
\label{tab:line_properties_with_lines}}

\tablehead{
\colhead{Region} &
\colhead{Component} &
\colhead{Line} &
\colhead{FWHM} &
\colhead{Centroid} &
\colhead{Intensity} \\
&
&
&
\colhead{(km s$^{-1}$)} &
\colhead{(\AA)} &
\colhead{(a.u.)}
}

\startdata
&  & [\ion{S}{2}]$\lambda6716$ & $78.0^{+15.8}_{-19.4}$ & $6829.4^{+0.2}_{-0.2}$ & $0.6^{+0.3}_{-0.2}$ \\
&  & [\ion{S}{2}]$\lambda6731$ & $78.0^{+15.8}_{-19.4}$ & $6844.7^{+0.2}_{-0.2}$ & $0.4^{+0.2}_{-0.2}$ \\
& Narrow & [\ion{N}{2}]$\lambda6548$ & $42.3^{+5.3}_{-4.4}$ & $6659.2^{+0.1}_{-0.1}$ & $0.7^{+0.1}_{-0.1}$ \\
&  & H$\alpha$ & $59.2^{+4.3}_{-3.9}$ & $6674.2^{+0.1}_{-0.1}$ & $5.4^{+0.5}_{-0.5}$ \\
&  & [\ion{N}{2}]$\lambda6584$ & $42.3^{+5.3}_{-4.4}$ & $6695.2^{+0.1}_{-0.1}$ & $2.0^{+0.3}_{-0.3}$ \\
\cline{2-6}
&  & [\ion{S}{2}]$\lambda6716$ & $223.2^{+17.3}_{-33.9}$ & $6831.8^{+0.2}_{-0.2}$ & $5.7^{+0.7}_{-1.5}$ \\
&  & [\ion{S}{2}]$\lambda6731$ & $223.2^{+17.3}_{-33.9}$ & $6847.1^{+0.2}_{-0.2}$ & $3.6^{+0.7}_{-0.9}$ \\
Region 1 & Intermediate & [\ion{N}{2}]$\lambda6548$ & $188.5^{+12.2}_{-12.8}$ & $6661.0^{+0.1}_{-0.1}$ & $4.3^{+0.6}_{-0.6}$ \\
&  & H$\alpha$ & $211.1^{+7.9}_{-13.4}$ & $6676.0^{+0.1}_{-0.1}$ & $21.7^{+1.9}_{-3.1}$ \\
&  & [\ion{N}{2}]$\lambda6584$ & $188.5^{+12.2}_{-12.8}$ & $6697.0^{+0.1}_{-0.1}$ & $13.0^{+1.9}_{-1.9}$ \\
\cline{2-6}
&  & [\ion{S}{2}]$\lambda6716$ & $489.1^{+76.2}_{-49.7}$ & $6830.3^{+0.9}_{-0.8}$ & $3.0^{+1.7}_{-0.7}$ \\
&  & [\ion{S}{2}]$\lambda6731$ & $489.1^{+76.2}_{-49.7}$ & $6845.6^{+0.9}_{-0.8}$ & $5.1^{+0.8}_{-0.7}$ \\
& Broad & [\ion{N}{2}]$\lambda6548$ & $481.6^{+38.6}_{-31.8}$ & $6660.4^{+0.2}_{-0.2}$ & $7.6^{+0.6}_{-0.5}$ \\
&  & H$\alpha$ & $488.9^{+38.8}_{-39.4}$ & $6675.4^{+0.2}_{-0.2}$ & $21.6^{+2.7}_{-1.2}$ \\
&  & [\ion{N}{2}]$\lambda6584$ & $481.6^{+38.6}_{-31.8}$ & $6696.4^{+0.2}_{-0.2}$ & $22.9^{+1.7}_{-1.6}$ \\
\hline
&  & [\ion{N}{2}]$\lambda6548$ & $240.5^{+25.4}_{-21.5}$ & $6658.6^{+0.6}_{-0.7}$ & $2.2^{+1.3}_{-1.2}$ \\
Region 2 & Single & H$\alpha$ & $278.5^{+23.4}_{-20.6}$ & $6675.4^{+0.2}_{-0.2}$ & $9.2^{+0.6}_{-0.6}$ \\
&  & [\ion{N}{2}]$\lambda6584$ & $240.5^{+25.4}_{-21.5}$ & $6696.6^{+0.3}_{-0.3}$ & $6.5^{+3.8}_{-3.7}$ \\
\hline
&  & [\ion{N}{2}]$\lambda6548$ & $203.2^{+16.1}_{-14.2}$ & $6660.4^{+0.5}_{-0.5}$ & $2.2^{+1.2}_{-1.2}$ \\
Region 3 & Single & H$\alpha$ & $183.6^{+19.9}_{-16.7}$ & $6675.6^{+0.2}_{-0.2}$ & $9.9^{+0.8}_{-0.8}$ \\
&  & [\ion{N}{2}]$\lambda6584$ & $203.2^{+16.1}_{-14.2}$ & $6696.9^{+0.2}_{-0.2}$ & $6.6^{+3.6}_{-3.7}$ \\
\enddata
\end{deluxetable*}

As mentioned in the previous section, we only observed emission lines associated with the filaments surrounding NGC 1275 in three of the five pointings targeted by ESPaDOnS. We therefore applied a dedicated fitting procedure to extract the intensity, velocity and velocity dispersion of the main emission lines. Thanks to the extremely high spectral resolution of the ESPaDOnS instrument, we were also able to fit multiple emission line components for these pointings allowing us to probe for the first time the potential superposition of filaments.

We will present here our methodology for the fitting procedure used to analyze the data. First, we extracted the wavelengths and spectral information from the ESPaDOnS data. Afterwards, we used the \texttt{sherpa} software (\citealt{2001SPIE.4477...76F}, \citealt{2007ASPC..376..543D}, \citealt{2024ApJS..274...43S}) to derive a preliminary and statistical fitting of the main emission lines. For this, we specified the number of emission components as Gaussian functions described by the amplitude, wavelength position of the emission peak and full-widths at half maximum (FWHM), allowing us to derive the intensity, velocity and velocity dispersion of each of the emission components. 

To remain physically sound, we tied the velocity of the \nii\text{} emission line doublet to the H$\alpha$ emission line velocity according to their determined wavelengths theoretical separation and specified that the amplitude of the \nii$\lambda6548$ emission line corresponds to one third of the \nii$\lambda6584$ emission line amplitude. Additionally, when multiple emission components were considered, we tied their respective FWHM for the \nii\text{} emission line doublet on one hand and the \sii\text{} emission line doublet on the other, while keeping the FWHM of the H$\alpha$'s emission components independent, owing to their different atomic masses and thermal broadening.
The velocity of the \sii$\lambda6731$ emission line was also fixed to the one of \sii$\lambda6716$ according to their theoretical wavelengths separation.

Following this, we used specific values of amplitude, position and FWHM for the H$\alpha$ emission line components to act as inputs for \texttt{sherpa} to perform the fitting of the emission lines. These input parameters were chosen as initial guesses based on visual inspection of the spectra. Moreover, the \texttt{sherpa} fitting used here apply a chi-squared statistic as well as a NelderMead optimization method and uses covariance as estimation method for the errors on the fit. Finally, thanks to the number of degrees of liberty and fit statistic, we are able to run statistical tests to estimate and determine which model and number of emission components best fit the ESPaDOnS emission spectra using the \texttt{calc\_ftest} and \texttt{calc\_mlr} functions from the \texttt{sherpa.utils} package. Applying these functions on Region 1 spectral fit, using the number of degrees of freedom and the final fit statistic, showed that the use of three emission components led to the lowest \texttt{ftest} and \texttt{mlr} scores compared to the use of simply two emission components. Per these statistical functions, this indicate that three emission components statistically better fit our emission spectra compared to only two. We also ran similar tests on the spectra of Regions 2 and 3, but this time investigating whether one or two emission components would lead to better statistical fits of the spectra. In both cases, a single emission component model was preferred leading to lower \texttt{ftest} and \texttt{mlr} scores overall. We also used the \texttt{dynesty} module (\citealt{2020MNRAS.493.3132S}, \citealt{2004AIPC..735..395S}) to perform a bayesian fit with initial values around our best-fitting results, which also favoured a three-component fitting model for Region 1.

To further deepen our fitting analysis, we used a Markov-Chain Monte-Carlo (MCMC) algorithm with the \texttt{emcee} (\citealt{2013PASP..125..306F}) module to test the parameters we obtained with the \texttt{sherpa} software. This offers us an additional degree of confidence on our fits as the algorithm walks through the parameter space and tests a large number of sample values, ensuring that the fitting converges to the best set of parameters. We therefore explored a significant sample space around the best fit values we previously obtained with \texttt{sherpa}. We list in Appendix \ref{annex:mcmc_priors} our defined priors for each of the parameters explored during the fitting procedure. The MCMC algorithm allows us to run a large number of walkers, here 2000, where each one will explore parameter values for the amplitude, FWHM and position of each lines based on a normal distribution of $1e^{-3}$ around the best fit values obtained previously, leading to an autocorrelation value of $129.6$, representing the number of steps needed for a walker to produce an independent sample, with a burn-in of $250$ steps. Afterwards, we are able to obtain the resulting distribution of parameters as well as the values that best fit the corresponding spectra. We display these fitted spectra in Figure \ref{fig:fitted_spectra} and 
the best fitting results of our MCMC algorithm for all three pointings in Table \ref{tab:line_properties_with_lines}.

Following this fitting procedure, we are able to derive emission line ratios for each of the detected emission components. Since we are limited in the number of emission lines observed in the ESPaDOnS data, we can only determine the following emission line ratios, namely: \nii$\lambda6548$/H$\alpha$, \sii$\lambda6716$/H$\alpha$ and \sii$\lambda6716$/\sii$\lambda6731$. To do this, we simply perform the ratio of the amplitudes of the emission lines displayed in Table \ref{tab:line_properties_with_lines}. All the emission line ratios obtained are thus listed in Table \ref{tab:line_ratios}.

\begin{deluxetable*}{l l l r}
\tablecaption{Emission Line Ratios and $3\sigma$ Errors for Each Kinematic Component\label{tab:line_ratios}}

\tablehead{
\colhead{Region} &
\colhead{Component} &
\colhead{Line Ratio} &
\colhead{Value}
}

\startdata
&  &
[\ion{S}{2}]$\lambda6716$ + [\ion{S}{2}]$\lambda6731$ / H$\alpha$ &
$0.2^{+0.1}_{-0.1}$ \\
 & Narrow &
[\ion{S}{2}]$\lambda6716$ / [\ion{S}{2}]$\lambda6731$ &
$1.5^{+1.0}_{-0.9}$ \\
&  &
[\ion{N}{2}]$\lambda6584$ / H$\alpha$ &
$0.4^{+0.1}_{-0.1}$ \\
\cline{2-4}
&  &
[\ion{S}{2}]$\lambda6716$ + [\ion{S}{2}]$\lambda6731$ / H$\alpha$ &
$0.4^{+0.1}_{-0.1}$ \\
Region 1 & Intermediate &
[\ion{S}{2}]$\lambda6716$ / [\ion{S}{2}]$\lambda6731$ &
$1.6^{+0.4}_{-0.5}$ \\
&  &
[\ion{N}{2}]$\lambda6584$ / H$\alpha$ &
$0.6^{+0.1}_{-0.1}$ \\
\cline{2-4}
&  &
[\ion{S}{2}]$\lambda6716$ + [\ion{S}{2}]$\lambda6731$ / H$\alpha$ &
$0.4^{+0.1}_{-0.1}$ \\
& Broad &
[\ion{S}{2}]$\lambda6716$ / [\ion{S}{2}]$\lambda6731$ &
$0.6^{+0.3}_{-0.2}$ \\
&  &
[\ion{N}{2}]$\lambda6584$ / H$\alpha$ &
$1.0^{+0.1}_{-0.1}$ \\
\cline{2-4}
Region 2 & Single &
[\ion{N}{2}]$\lambda6584$ / H$\alpha$ &
$0.7^{+0.1}_{-0.1}$ \\
\hline
Region 3 & Single & [\ion{N}{2}]$\lambda6584$ / H$\alpha$ &
$0.7^{+0.1}_{-0.1}$ \\
\enddata
\end{deluxetable*}



\subsection{Polarimetric Data Analysis}

\begin{figure*}[hbt!]
\begin{subfigure}{0.48\linewidth}
  \centering
  \includegraphics[width=\linewidth]{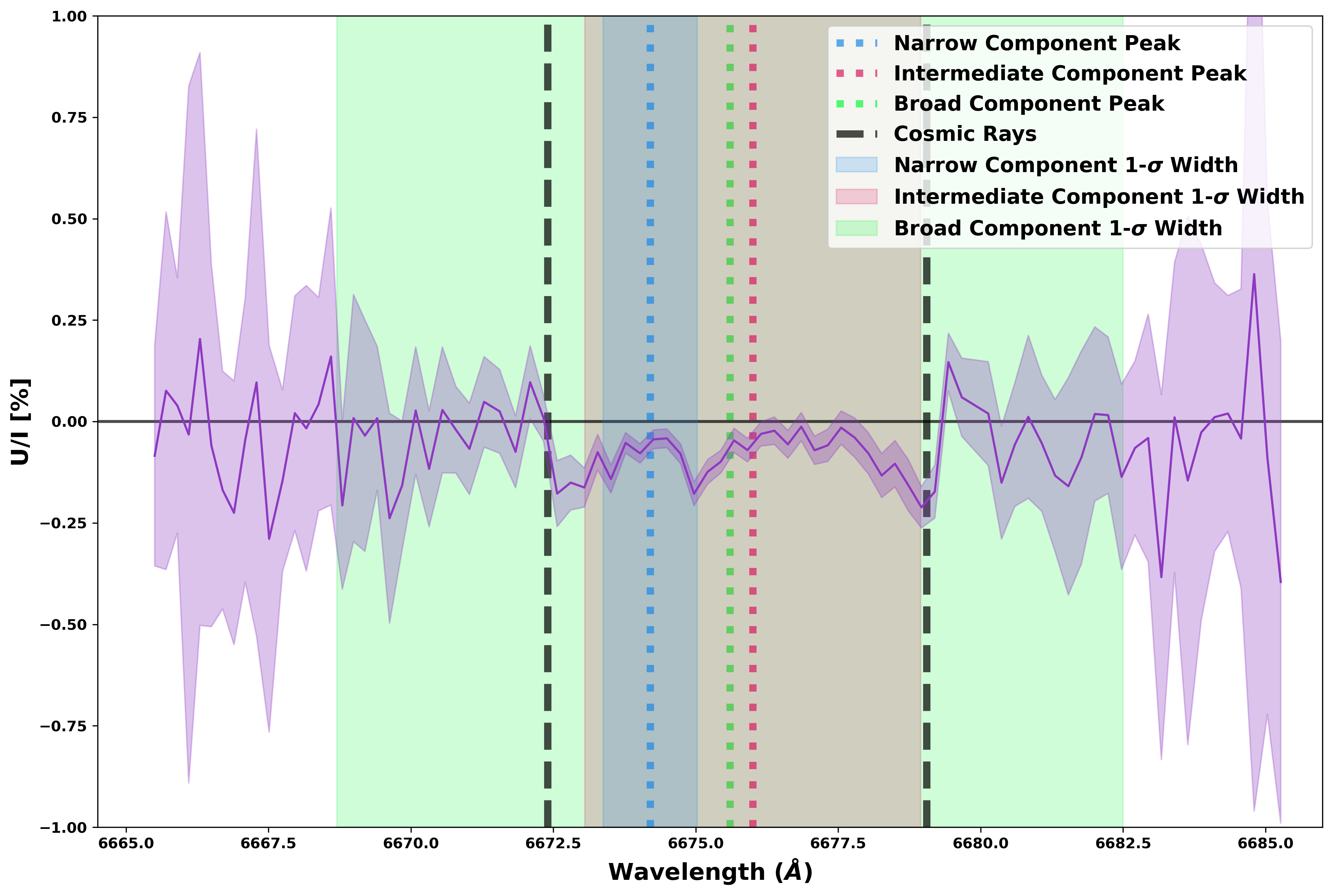}
  \label{flux}
\end{subfigure} 
\medskip
~ 
\begin{subfigure}{0.48\linewidth}
  \centering
  \includegraphics[width=\linewidth]{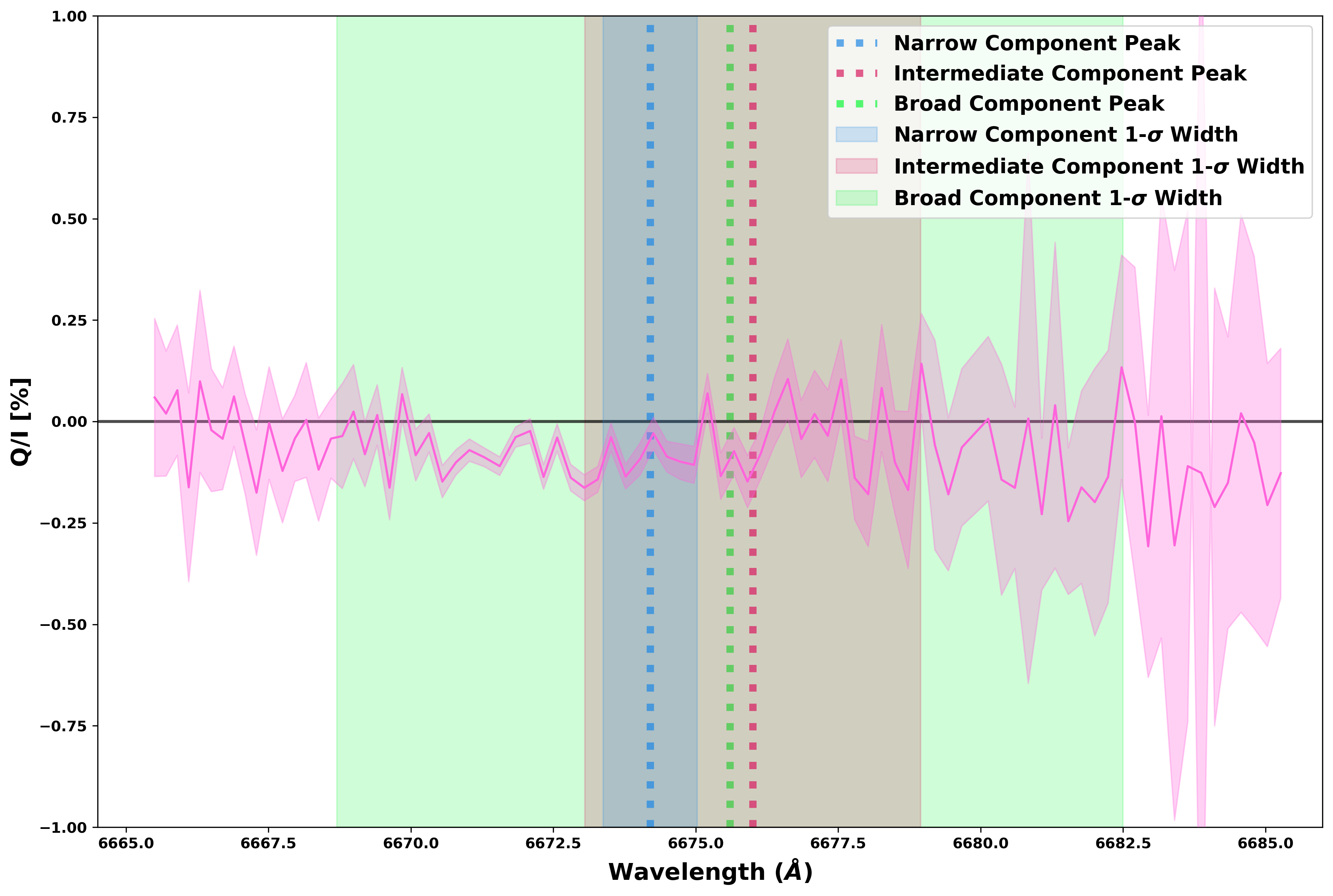}
  \label{degree}
\end{subfigure}
\caption{Left: ESPaDOnS normalized and binned Stokes U parameter and associated error centered on the H$\alpha$ emission line position for the first region. We display the 1-$\sigma$ widths of the emission components as shaded areas and the peaks of the three emission line components as colour-coded dotted lines derived from Table \ref{tab:line_properties_with_lines}, as well as the positions of the observed cosmic rays as black dashed line. Right: ESPaDOnS normalized and binned Stokes Q parameter and associated error centered on the H$\alpha$ emission line position for the first region. We display the 1-$\sigma$ widths of the emission components as shaded areas and the peaks of the three emission line components as colour-coded dotted lines derived from Table \ref{tab:line_properties_with_lines}.}
\label{fig:Stokes_parameters}
\end{figure*}

\begin{figure*}[hbt!]
\begin{subfigure}{\linewidth}
  \centering
  \includegraphics[width=0.5\linewidth]{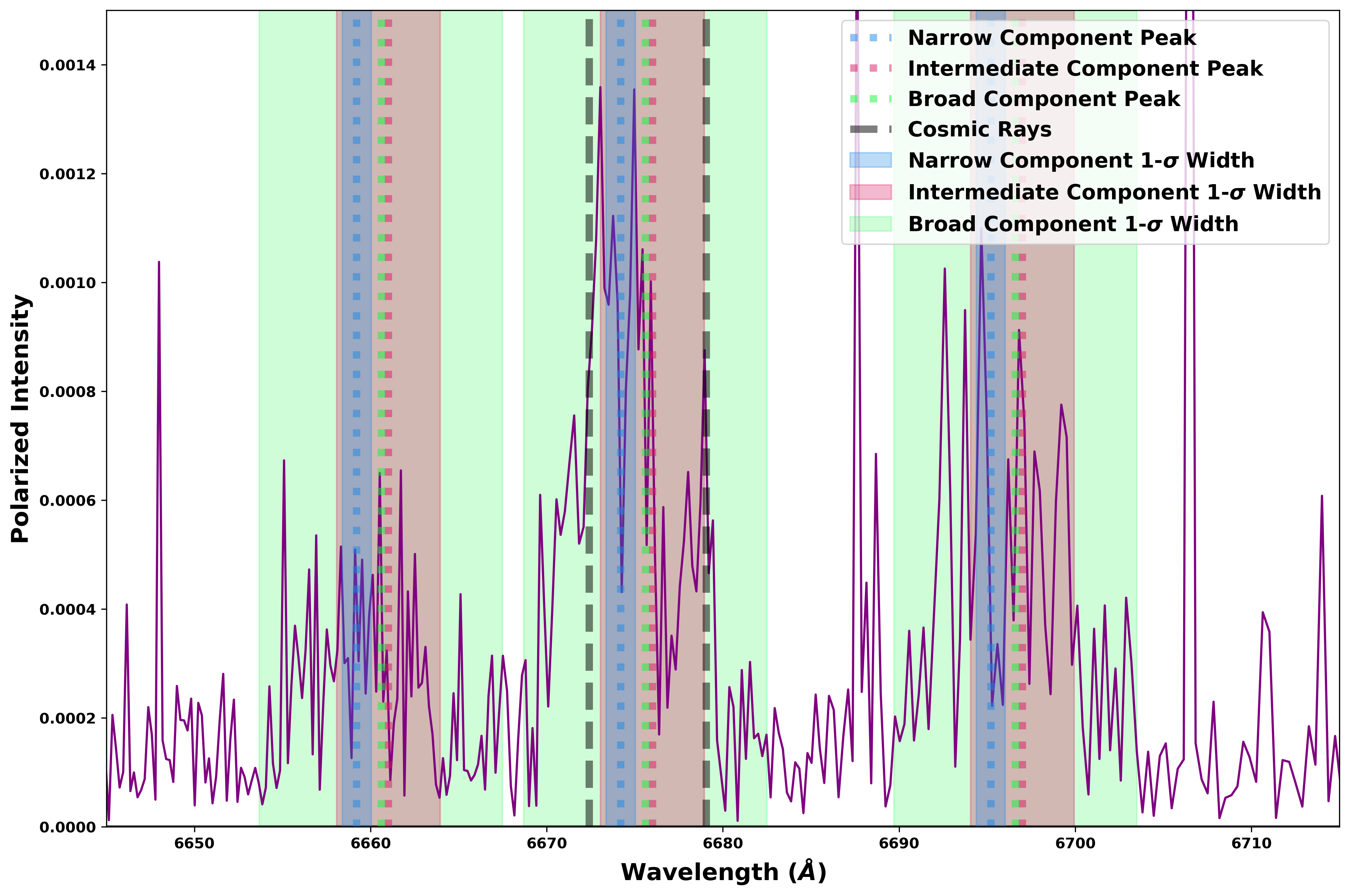}
  \label{flux}
\end{subfigure} 
\medskip
~ 
\begin{subfigure}{0.5\linewidth}
  \centering
  \includegraphics[width=\linewidth]{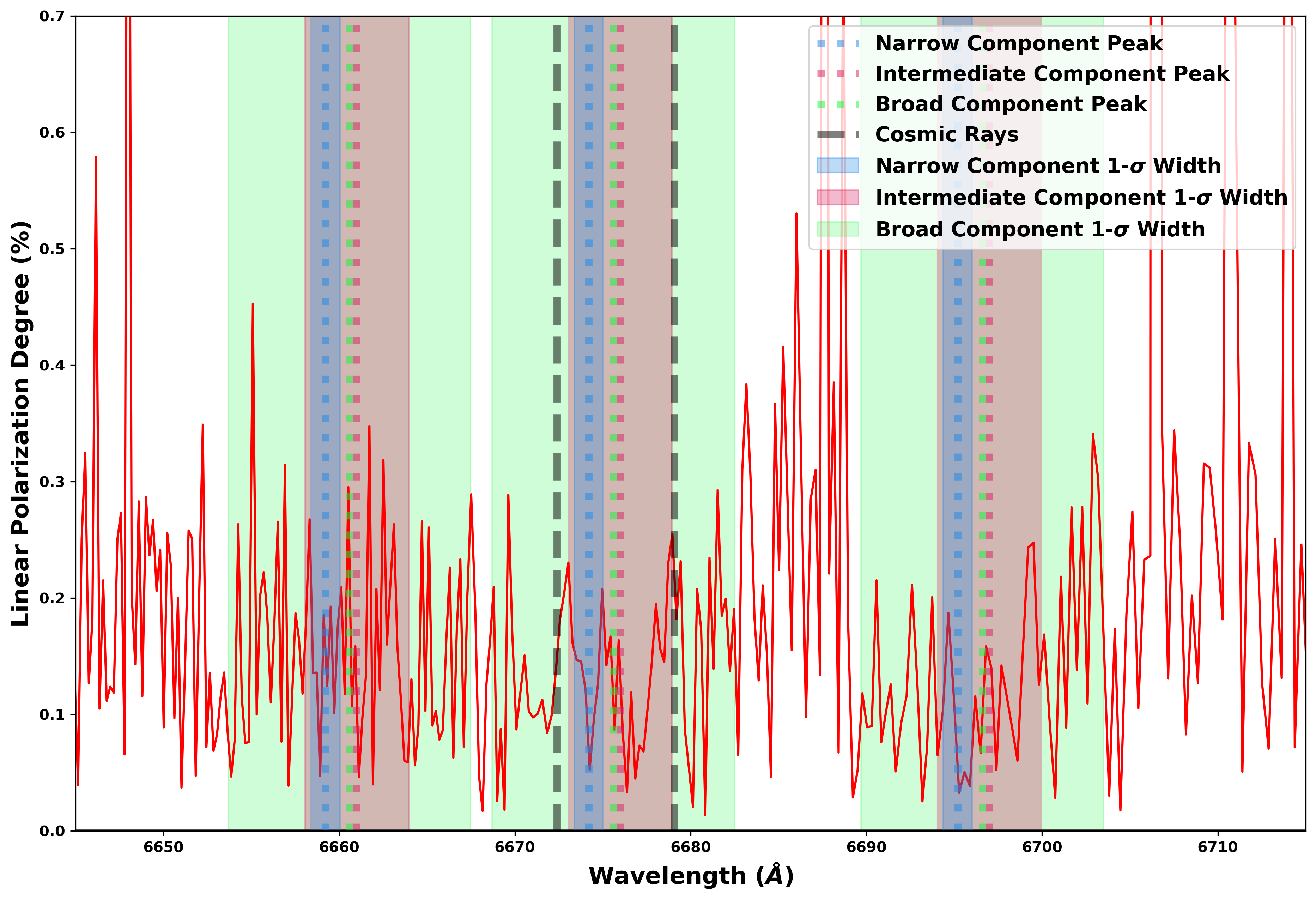}
  \label{degree}
\end{subfigure}
\begin{subfigure}{0.5\linewidth}
  \centering
  \includegraphics[width=\linewidth]{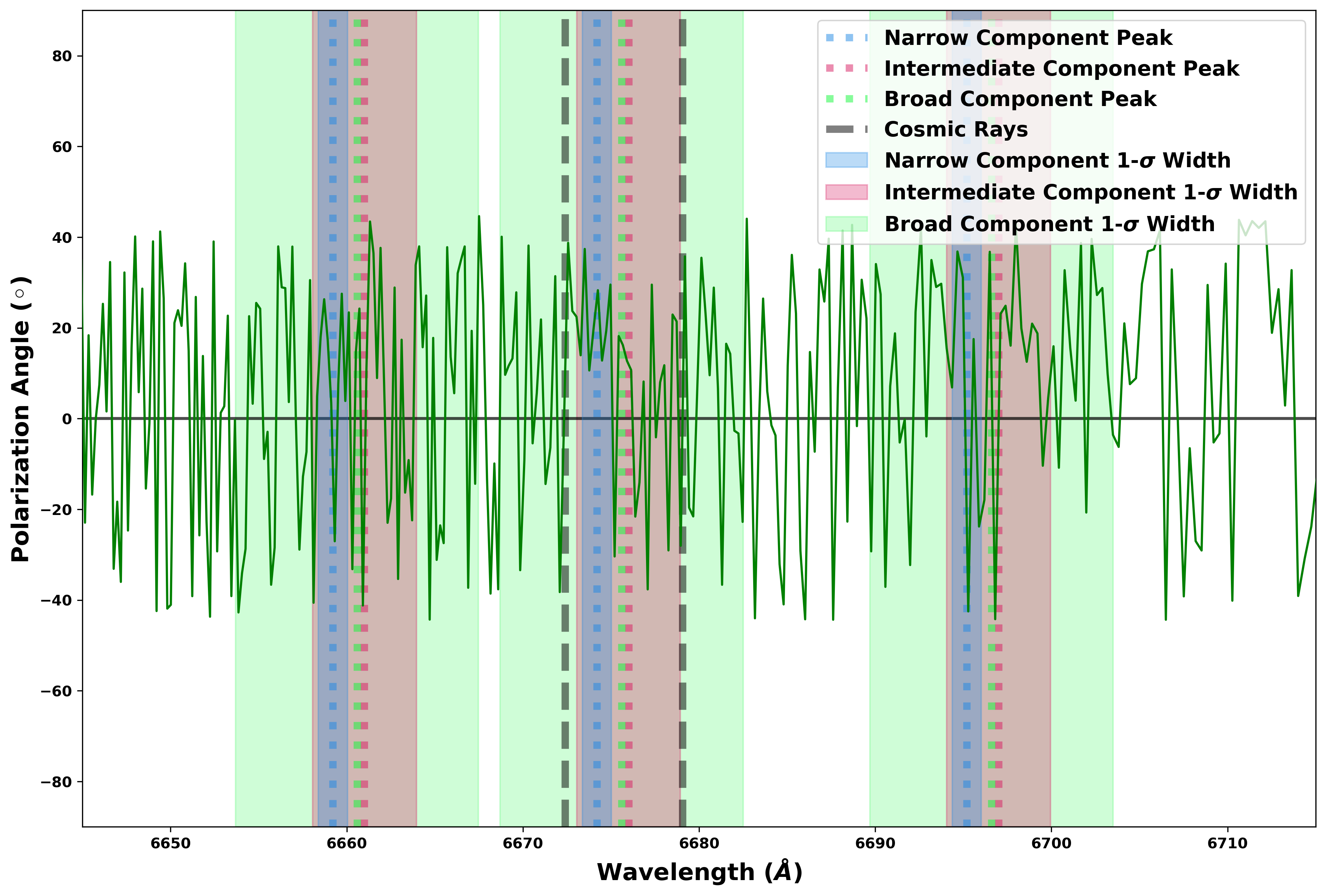}
  \label{angle}
\end{subfigure}
\caption{Top: ESPaDOnS measured polarized intensity for the first region derived from the normalized and binned Stokes Q and U parameters and their respective intensities $I_q$ and $I_u$ according to Equation \ref{eq:polarized_intensity}. We display the 1-$\sigma$ widths of each emission component as colour-coded shaded areas and the peaks of the three emission line components as colour-coded dotted lines derived from Table \ref{tab:line_properties_with_lines} as well as the positions of the observed cosmic rays as black dashed line. Bottom Left: ESPaDOnS measured linear polarization degree for the first region derived from the normalized and binned Stokes Q and U parameters according to Equation \ref{eq:polarization_degree}. We display the 1-$\sigma$ widths of each emission component as colour-coded shaded areas and the peaks of the three emission line components as colour-coded dotted lines derived from Table \ref{tab:line_properties_with_lines} as well as the positions of the observed cosmic rays as black dashed line. Bottom Right: ESPaDOnS measured polarization angle for the first region derived from the normalized and binned Stokes Q and U parameters according to Equation \ref{eq:angle_polarization}. We display the 1-$\sigma$ widths of each emission component as colour-coded shaded areas and the peaks of the three emission line components as colour-coded dotted lines derived from Table \ref{tab:line_properties_with_lines} as well as the positions of the observed cosmic rays as black dashed line.}
\label{fig:polarization_degree_and_angle}
\end{figure*}

After carefully investigating the polarimetric spectra obtained with ESPaDOnS of the three pointings studied previously, we noticed that for Region 1 displaying the highest SNR, some tentative features associated with the brightest emission line, namely H$\alpha$, appeared visible in the Stokes Q and U polarimetric spectra. However, the ESPaDOnS polarimetric data obtained during these observations showed a relatively high level of noise complicating the analysis and detection of such features. No clear detection of polarimetric features was however observed in Regions 2 and 3, most likely due to the poor SNR obtained for these pointings.

Past studies of ESPaDOnS' polarimetric data displaying high levels of noise have been carried out and allowed the development of mathematical tools to properly perform a variable binning depending on the relative error of the data (see \citealt{de2013search, de2014searching}). Applying these methods to our polarimetric observations offers us a much improved detection of the features mentioned previously as well as a reduction of the overall noise level.

We will therefore describe below the procedure used to obtain the variable binning of the polarimetric data. First, a relative error threshold associated with our polarimetric observations is defined to determine the level of binning needed. Investigating our polarimetric data and its error, we established that threshold at $0.00016$ to obtain an improved SNR for the polarimetric data. Following this, the algorithm will incrementally bin the polarimetric data to reach the error threshold across the entire spectrum. This result in a variably binned polarimetric spectrum for the Stokes U and Q parameters allowing us to study the optical linear polarization of the filaments surrounding NGC 1275. 

We thus display in Figure \ref{fig:Stokes_parameters} left and right panels the features visible in the normalized and binned Stokes U and Q respectively associated with the most prominent H$\alpha$ emission line for Region 1. Additionally, we indicate the position of the peaks and 1-$\sigma$ widths of the three detected emission components based on our fitting results obtained in Section \ref{sec:fitting} as colour-coded dotted lines and shaded areas respectively. As mentioned in Section \ref{sec:obs}, we also indicate for the binned and normalized U Stokes parameter the position of the detected cosmic rays as black dashed lines. 

Based on these binned spectrum, we are able to calculate the linear polarization degree, polarized flux as well as the angle of polarization using the following formulas:

\begin{equation}
\label{eq:polarization_degree}
    \Pi_{lin} = \sqrt{u^2 + q^2},
\end{equation}

\begin{equation}
\label{eq:polarized_intensity}
    I_{pol} = \Pi_{lin} \times \sqrt{I_u^2 +I_q^2},
\end{equation}

\begin{equation}
\label{eq:angle_polarization}
    \phi_{lin} = \frac{1}{2}\text{arctan}\left(\frac{u}{q}\right),
\end{equation}

where q and u correspond to the normalized Q and U Stokes parameters, respectively: $100\times Q/I_q$ and $100\times U/I_u$. We display our resulting measurements for each of these parameters, in Figure \ref{fig:polarization_degree_and_angle}. Similarly, we indicate the peaks and widths of the three detected emission components based on our fitting results obtained in Section \ref{sec:fitting} as colour-coded dotted lines and shaded areas.

We discuss the interpretation of these various polarimetric features and results in Section \ref{sec:polarization}.



\section{Results and Discussion} \label{sec:results}

\subsection{Emission Line Intensity Ratios} \label{sec:ratios}

We first discuss the emission line ratios obtained from the fitting of the main emission lines as described in Section \ref{sec:fitting} and shown in Table \ref{tab:line_ratios}. As explained previously, we are limited in the number of emission line ratios obtainable through the ESPaDOnS spectral data, since only a limited set of emission lines are detectable in the three pointings. Nevertheless, from Region 1 specifically, we are able to determine three main emission line ratios namely: \sii\text{} by H$\alpha$, \sii$\lambda6716$ by \sii$\lambda6731$ and \nii$\lambda6584$ by H$\alpha$. For Regions 2 and 3 however, we are only able to determine the \nii$\lambda6584$ by H$\alpha$ emission line ratio. These various ratios gives us information on the ionization mechanisms (see \citealt{kewley2019understanding}) as well as the electron density (see \citealt{1989agna.book.....O}) of the optical gas composing the filaments surrounding NGC 1275.

\subsubsection{Region 1}

\begin{figure}
    \centering
    \includegraphics[width=\linewidth]{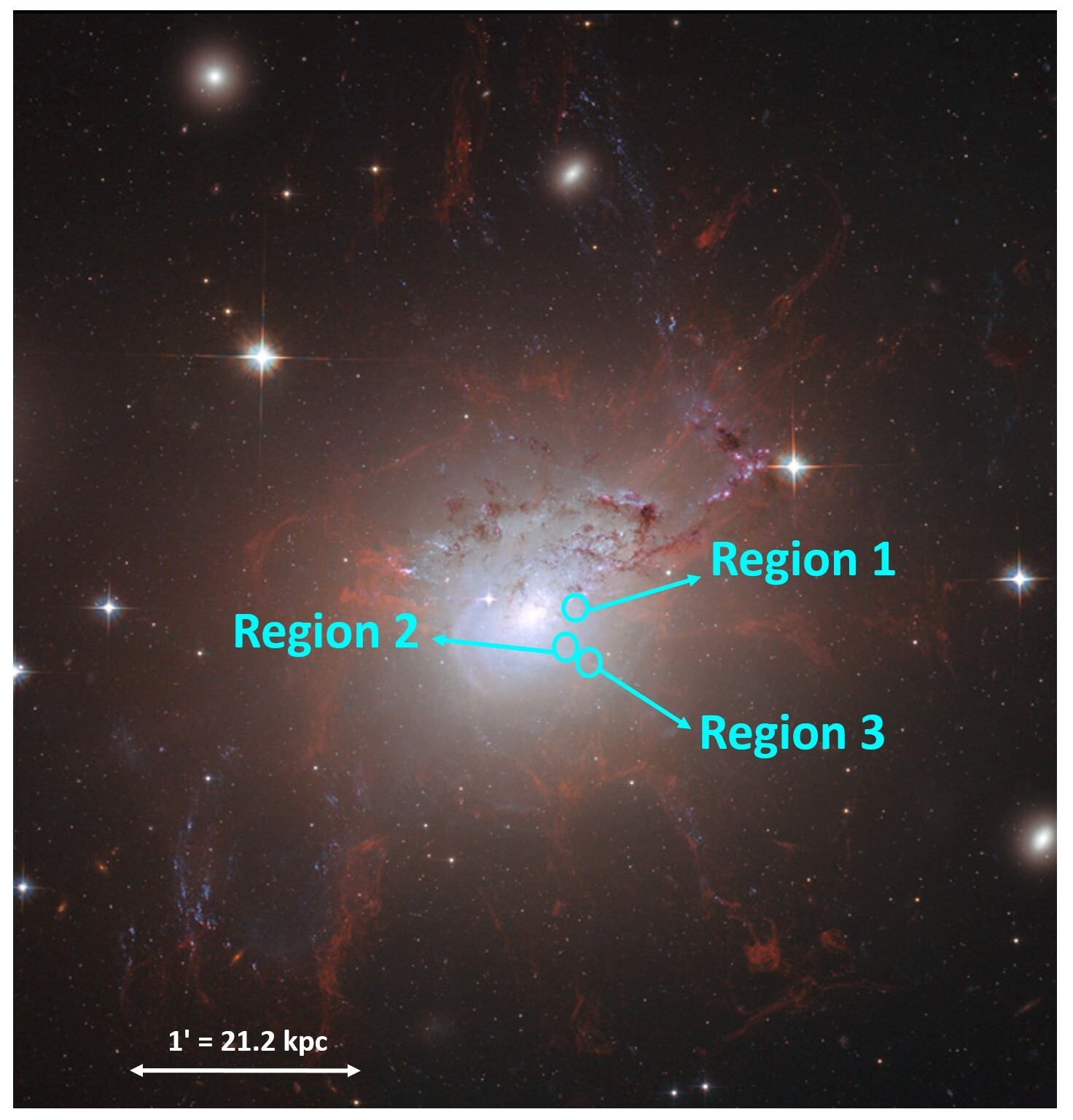}
    \caption{Location of the three ESPaDOnS pointings displaying emission signal observing key locations within the filamentary nebula surrounding NGC 1275 overplotted on composite HST observations (HST GO 10546 - \citealt{fabian_magnetic_2008}), where the blue component traces the UV emission mainly coming from younger stars. The presence of dust from the foreground galaxy called High-Velocity System is also visible in absorption.}
    \label{fig:hst_pointings}
\end{figure}

We will first start by examining the emission line ratios of Region 1, whose pointing displays multiple emission components. Thanks to its relatively high SNR of $\sim 30$, we are able to detect and determine the \sii\text{} by H$\alpha$, \sii$\lambda6716$ by \sii$\lambda6731$ and \nii$\lambda6584$ by H$\alpha$ emission line ratio for three emission line components as displayed in the first row of Table \ref{tab:line_ratios}.

Focusing first on the \nii$\lambda6584$ by H$\alpha$ ratio, it informs us on the predominance of strong versus soft sources of ionization. In that sense, a relatively low ratio (meaning H$\alpha$ flux $\gg$ \nii$\lambda6583$ flux) could indicate the presence of young stars or a softer ionizing source (see \citealt{canning2010star}, \citealt{canning_filamentary_2014} and Figure \ref{fig:hst_pointings}, where we display the position of our regions over composite HST observations showing in blue the presence of UV emission associated with young stars), while a relatively high ratio (meaning H$\alpha$ flux $\leq$  \nii$\lambda6583$ flux) could indicate the presence of a hard ionizing source such as an AGN or energetic ICM particles. Interestingly, from our measurements of the \nii$\lambda6584$ by H$\alpha$ ratio for the first pointing, we can see that the narrower emission component displays a low ratio of $\sim0.4$, while the intermediate and broader components indicate larger ratio of respectively $\sim0.6$ and $\sim1.0$. This indicates that the broadest component is most likely ionised by a harder source, presumably the AGN as the central galaxy is located $\sim 6.8''$ or $\sim 2.5$ kpc away from the ESPaDOnS' pointing, or possibly through a local hard excitation source, while the narrower component could seemingly be ionised by a softer source. Another ionization mechanism explaining the energetics of optical filaments around BCGs could be through the collisional excitation coming from the surrounding hot ICM (see \citealt{ferland_collisional_2009}) leading to a \nii$\lambda6548$ / H$\alpha$ emission line ratio of $\sim 0.3$, which is close to the result we observe for the narrowest component. The intermediate component, on the other hand, indicates a moderately high \nii$\lambda6548$ / H$\alpha$ ratio at $\sim 0.7$ possibly hinting at a combination of ionizing sources but likely influenced by a harder source such as the central AGN.

Additionally, the ratio of \sii\text{} by H$\alpha$(6563\AA) can be used as a confirmation value for various ionization mechanisms (\citealt{1989agna.book.....O}). As can be seen in Table \ref{tab:line_ratios} for the first pointing, most of the \sii\text{} by H$\alpha$(6563\AA) ratio values we obtain for our emission components are extremely low (around $\sim 0.2$ for the narrowest component and at $\sim 0.4$ for the intermediate and broad components respectively). These values appear overall low and would prevent us from offering a clear conclusion regarding a potential ionization by collisional excitation, since these would lead to a ratio superior to $1.0$ (see \citealt{ferland_collisional_2009}), or through the activity of the AGN. Moreover, since we do not clearly detect the presence of [O III] or H$\beta$ emission within the ESPaDOnS spectra, due to either no emission or the presence of instrumental artifacts affecting the data (see Appendix \ref{annex:hbeta_oiii}, where we display the ESPaDOnS spectra at the position of the [O III] and H$\beta$ emission lines), we are unable to perform a complete Baldwin, Phillips \& Terlevich, or BPT diagram (e.g. \citealt{baldwin_classification_1981}) allowing us to more precisely derive the ionizing source of the detected emission components. We do note however, that [O III] and H$\beta$ emission lines are reported and detected in prior SITELLE observations of the filamentary nebula surrounding NGC 1275 (see \citealt{rhea2025multiwavelength}). 

Finally, the \sii$\lambda6716$ by \sii$\lambda6731$ emission line ratio can inform us on the electron density of the optical gas composing the filaments surrounding NGC 1275 at a temperature of $\sim 10^4$ K (e.g. \citealt{1989agna.book.....O}). Collisional deexcitation is a commonly used mechanism to determine the electron density of such gaseous structure. By considering the emission lines of ions with similar excitation energy but different energy levels, their relative excitation rate will only depend on the ratio of collisional strengths. Additionally, specific emission lines possess energy levels with different radiative transition probabilities or collisional deexcitation rates, thus, calculating their respective intensity ratios will give us a measure of the electron density of the gas. An instance of such line ratio is \sii$\lambda6716$ by \sii$\lambda6731$ (see \citealt{1989agna.book.....O}).


The ratio values to electron density relationship was determined by \cite{1989agna.book.....O}, which also demonstrated the presence of upper ($\sim 1.3$) and lower ($\sim 0.4$) ratio values above and below which electron density measurements cannot be derived effectively. Nonetheless, \cite{proxauf2014upgrading} developped an updated mathematical relationship using improved atomic data to derive the electron density of optical gas at a temperature of $T=10000$ K from the \sii\text{} emission lines ratio.


This updated formula was thus used to calculate the electron density for the first pointing's emission components using our ratio values obtained previously, following:


\begin{equation}
    \begin{split}
        \text{log}(n_e[\text{cm}^{-3}]) = &\text{ } 0.0543 \text{ tan}(-3.0553\text{ }R + 2.8506)\\
        & + 6.98 - 10.6905\text{ }R\\
        & +9.9186\text{ }R^2 - 3.5442\text{ }R^3,
    \end{split}
    \label{eq:density_proxauf}
\end{equation} 

where R corresponds to the measured ratio of \sii$\lambda6716$ and \sii$\lambda6731$. As mentioned previously, values of electron densities are however strongly bounded by threshold values below and above which the resulting values are unreliable. We can thus see that the narrow and intermediate emission component, with respective ratios of $1.5\pm1.0$ and $1.6\pm0.5$, both fall outside the bound of the relationship and would most likely indicate lower electron densities. Nevertheless, we can determine the values of electron density for the broad emission component leading to a value of $2875.9 ^{}_{-2223.6} \text{ cm}^{-3}$, where the upper error is unconstrained due to the lower error bound of the ratio being near the diagnostic’s saturation limit (see \citealt{proxauf2014upgrading}).

These results indicate a tentative trend of increasing electron density (or decreasing \sii\text{} emission line ratio) from the narrowest component to the broadest, possibly implying a preferential origin for the filaments and the accumulation of cooler gas, which, when put in conjunction with the ionization trends described above, could also tentatively inform us on the potential superposition of filamentary structures, which will be discussed in Section \ref{sec:superposition}. Interestingly, a recent analysis of the \sii\text{} emission line doublet from the SITELLE observations revealed a more complex picture of density measurements sharing a few similarities (\citealt{Vigneron_2026}). First, due to a poor SNR for the outer filaments, the authors considered large swathes of filaments to obtain a strong enough signal and properly measure an overall density. This, however, leads to a loss of the density spatial scale across the filaments since only a single density value is considered for an entire filament, leading to higher density measurement overall close to $\sim 350.3\text{ cm}^{-3}$ on average for outer filaments. However, for the inner filaments, the authors were able to perform a radial analysis of the \sii\text{} ratio due to a significantly higher SNR. These measurements, indicate relatively low and radially stable density close to the threshold value of $\sim 10^{2} \text{ cm}^{-3}$ for the inner filaments.

\subsubsection{Regions 2 and 3}

We can now consider the second and third ESPaDOnS regions for which we only have access to the \nii$\lambda6584$ by H$\alpha$ emission line ratio. From these, we can only determine a partial profile for the ionization source of the optical emission observed within these pointings. Indeed, for both pointings, we can see that the emission line ratios are of $0.70$ indicating a potential hard or mixed source of ionization. Ratio-wise, they would appear relatively similar to the intermediate component observed in the first pointing, while interestingly displaying similarly high or higher velocity dispersions (see Table \ref{tab:line_properties_with_lines}). Nevertheless, the absence of other detected emission lines within Regions 2 \& 3 prevents us from drawing a particular conclusion regarding the ionization model or density properties of the filaments.

\subsection{Superposition of Filaments and AGN Driven Outflow}
\label{sec:superposition}

We now discuss the main emission lines components detected within the ESPaDOnS spectral observations of the filaments surrounding NGC 1275 and how these informs us on the possibility of superimposed filamentary structures along the line of sight.

First, for Regions 2 \& 3 allowing us to only correctly fit a single component model, the significantly lower SNR prevents us from identifying multiple components, which suggests that improved observations could potentially reveal additional emission components or, otherwise, that these regions are indeed only composed of a single broader emission component. The current spectral observations prevents us from singling out one scenario over the other.

On the other hand, Region 1, located $\sim 6.8''$ away from the BCG, displays multiple emission components as seen in Figure \ref{fig:fitted_spectra} top panels and Table \ref{tab:line_properties_with_lines}. Interestingly, the intermediate and narrow emission components have velocity dispersions of $\sim 89.6$ km/s and $\sim 25.1$ km/s respectively for the H$\alpha$ emission line, which appear slightly lower than observed velocity dispersion measurements presented in \cite{vigneron2024high} for both the central more turbulent elongated structure near the BCG displaying an average velocity dispersion of $\sim 134$ km/s and the outer more quiescent filaments displaying on average a velocity dispersion of $\sim 44$ km/s. This could therefore suggest that the observed two components in the ESPaDOnS spectral data could represent a superposition of both emission structures detected in the high-spectral-resolution SITELLE observations but significantly more constrained due to the higher spectral resolution of the ESPaDOnS observations.

From the fitting results we have obtained with our MCMC algorithm as displayed in Table \ref{tab:line_properties_with_lines}, we are able to determine the relative velocity difference between each of the components and the corresponding velocity of the central BCG determined according to the redshift of $0.017284$ based on stellar velocities (Hitomi \citealt{hitomi_collaboration_atmospheric_2018}) and taking into account the barycentric velocity correction at the time of the observations, using \texttt{astropy} \texttt{radial\_velocity\_correction} function and established at -3.28 km/s. Thus, for the narrow emission component, we find that its line-of-sight velocity shift would correspond to $-87.0\substack{+1.1 \\ -1.1}$ km/s, indicating a strongly blueshifted emission. On the other hand, the intermediate and broad emission components indicate line-of-sight velocity shifts of $-6.7\substack{+4.2 \\ -4.0}$ km/s and $- 26.0\substack{+5.2 \\ -5.7}$ km/s respectively indicating slightly less blueshifted structures in comparison to the overall stars. Therefore, the ESPaDOnS spectroscopic observations appear to indicate a gradient of velocities between the various emission components giving a tentative idea of their possible superposition. Additionally, this detection of multiple emission components suggests a more complex kinematic picture for filamentary nebula surrounding BCGs. Indeed, the archival ESPaDOnS observations reveal multiple emitting structures within a single pointing of the filaments, namely with two emission components (narrow and broad) significantly offset velocity-wise from the stars, while the intermediate component appears to be closer to the stellar velocity. Past studies demonstrated that nebular gas, often considered as a single emitting structure, was mostly kinematically decoupled from the stellar population (e.g. \citealt{hamer_optical_2016}, \citealt{Gingras_2026}). However, ESPaDOnS' extremely high spectral resolution, shows a clear detection of multiple emission components along the line of sight each with separate velocity offsets from the stellar population allowing us to obtain a more comprehensive representation of the filamentary structure's kinematics in comparison to the stellar population.

Thus, when put in conjunction with the emission line intensity ratios obtained in Section \ref{sec:ratios}, we seem to obtain a clearer picture of the possible overlapping of these emission structures. Indeed, the strongly blueshifted ($-87.0\substack{+1.1 \\ -1.1}$ km/s) narrow emission component, displaying a possibly low electron density according to its \sii\text{} emission doublet ratio, as well as a less energetic ionizing source could most likely be an outer filament seen along the line of sight and located further away from the central BCG. On the other hand, the slightly blueshifted intermediate component ($-6.7\substack{+4.2 \\ -4.0}$ km/s), with a velocity dispersion of $\sim 82.1$ km/s, displaying a low electron density from its \sii\text{} emission doublet ratio and a more energetic ionizing source appears to correspond to the identified central elongated and turbulent structure as seen in \cite{vigneron2024high}, which could originate from a region of churned material under the effect of the currently forming radio bubbles inflated by the jets of the SMBH at the center of NGC 1275. Finally, the broadest, yet faintest, emission component with a velocity dispersion of $\sim 207.6$ km/s for the H$\alpha$ emission line appears blueshifted ($- 26.0\substack{+5.2 \\ -5.7}$ km/s) and displays a high electron density of $\sim 2875.9\text{ cm}^{-3}$ as well as an energetic ionizing source and could most likely correspond to the observed more turbulent central elongated structure as observed in past SITELLE observations (\citealt{vigneron2024high}) or be a detected outflow emission from the central AGN since the ESPaDOnS pointing for this spectroscopic measurement is located only $6.8''$ away from the position of the central BCG. 

Therefore, the extremely high spectral resolution ($\text{R}=68000$) of the ESPaDOnS observations appears to reveal for the first time a more comprehensive superposition of emission features associated with the filamentary nebula surrounding NGC 1275, while better constraining their velocity dispersions. Consequently, carrying out spectroscopic observations of similar spectral resolution of filaments located around BCGs of galaxy clusters could offer us the opportunity to better disentangle and understand their spatial arrangement, as well as explore the tentative presence of extremely narrow emission components associated with more distant quiescent filaments that could have been missed in previous spectroscopic mission with too low spectral resolutions. Hence, recent high-spectral-resolution ($\text{R}=56000$) Gemini South/GHOST spectroscopic observations of the optical filaments surrounding NGC 4696, BCG of the nearby Centaurus Cluster, also appear to display multiple emission components within several pointings across the filaments, some of which appear extremely narrow with velocity dispersions at or below $\sim 20$ km/s (Vigneron et al. in preparation).


\subsection{Polarization Measurements}
\label{sec:polarization}

We will now discuss the polarization features and results we observe in the spectropolarimetric ESPaDOnS observations of Region 1 within the filamentary nebula surrounding NGC 1275. 

\subsubsection{Polarization Constraints}
First, focusing on the measured polarized flux displayed in the top panel of Figure \ref{fig:polarization_degree_and_angle} for the H$\alpha$ and \nii\text{} complex, we can observe a significant polarized flux increase in the wings of the main emission lines, as visible in the colour-coded shaded areas corresponding to the $1-\sigma$ widths of each components, but also a significant drop at the peaks of the emission components within the emission lines, which are represented as colour-coded dotted lines in the top panels of Figure \ref{fig:polarization_degree_and_angle}. This result strongly suggests that the main emission lines are overall less polarized than the surrounding continuum, a result both observed in \cite{sparks2013polarization} and \cite{marin2025spectropolarimetry} when studying the optical polarized flux of similar structures.

Additionally, the extremely high spectral resolution observations of ESPaDOnS, as well as the relatively good SNR of the variably-binned polarimetric data allows us to characterize the polarization degree over the position of our H$\alpha$ and \nii\text{} complex. Indeed, as displayed in the bottom left panel of Figure \ref{fig:polarization_degree_and_angle}, we can see significant drops of the optical linear polarization degree at the position of the emission components peaks, represented by colour-coded dotted lines, while we observe increases in the wings of the emission lines, represented as colour-coded shaded regions. This reinforces our previous interpretation of the polarized flux as the overall polarization of the emission lines seem to decrease in comparison to the continuum, which has a polarization degree of $\sim 0.26\pm0.30\%$ on average. Nevertheless, the polarization degree remains at an average value of $\sim 0.15 \pm 0.08\%$ for H$\alpha$ and of $\sim 0.14\pm 0.12\%$ for \nii$\lambda6584$, which appear close to the reported value of $\leq 0.1 \%$ by \cite{sparks2013polarization} for other gaseous filaments. In their paper, the authors explored the polarimetric properties of various filamentary nebula surrounding BCGs, namely M87, NGC 4696, Hydra A and PKS0745-19, using slit-spectroscopy to discern sources of excitation within the nebular gas through measurements of their polarization degree. As mentioned previously, for each of the observed BCGs, low polarization degrees ($\sim 0.05 -  0.1 \%$) were observed, leading the authors to conclude that the H$\alpha$ emission lines could not be predominantly ionised by non-thermal particles from the ICM, since it would lead to linearly-polarized emission lines due to the particle flux being strongly anisotropic. The ESPaDOnS polarimetric observations presented here thus appear to comfort these past measurements. \cite{marin2025spectropolarimetry}, on the other hand, reported high linear polarization degrees of $\sim 1-2\%$ solely for the H$\alpha$ emission line of the luminous jets of NGC 1275, but requiring a significant spectral binning of $237$\AA.

Drawing our attention to the angle of polarization, presented in the bottom right panel of Figure \ref{fig:polarization_degree_and_angle}, it appears to remain stable around $5.3\pm25.7^\circ$ and independent of wavelength at the position of the H$\alpha$ and \nii\text{} complex, indicating that neither continuum nor filament's emission affect the orientation of the observed polarization.

Finally, considering the normalized u and q Stokes parameters centered on the most prominent H$\alpha$ emission line of Region 1 and displayed in Figure \ref{fig:Stokes_parameters}, we can identify clear non-negligible drops and features in both q and u. Interestingly, these features appear to be velocity-dependant as they do not align with one another at the position of the multiple emission components within H$\alpha$, yet again indicated by their peaks and $1-\sigma$ widths as colour-coded dotted lines and shaded regions respectively. Indeed, the clear drop in the u parameter appear more centered at $\sim 6676 \text{ \AA}$ at the central position of the H$\alpha$ emission line, while the slight drop in q appears more blueshifted in the wings of the H$\alpha$ emission line at around $\sim 6673 \text{ \AA}$. We also note, as previously mentioned in Section \ref{sec:data}, the presence of two cosmic rays generating significant drops and/or increases in the normalized u Stokes parameter, indicated as black dashed lines in Figure \ref{fig:Stokes_parameters}, which further complicates this polarization profile. We will thus explore, in the following subsection, tentative explanations to the observed linear polarization features within the optical emission line of the filaments surrounding NGC 1275.

\subsubsection{Potential Sources of Polarization}

Investigating now potential sources of polarization for the optically-emitting filaments surrounding NGC 1275, we will explore the possible roles of magnetic fields, dust scattering and superposition of varying polarization features. 

Indeed, previous studies on gaseous filamentary structures around BCGs argued that the presence of magnetic pressure dominating the thermal pressure allowed the filaments to remain stable preventing their collapse and significant star formation (see \citealt{fabian_relationship_2003}). However, such magnetic fields, estimated at $\sim 50 - 100\text{ } \mu$G (\citealt{fabian_magnetic_2008}, \citealt{werner2013nature}), could only be properly constrained through the analysis of circular polarization which wasn't effectively measured in the ESPaDOnS observations, through lack of V Stokes parameter measurements. Previous linear polarization analyses by \cite{sparks2013polarization} of spectropolarimetric observations of multiple filamentary nebula around the BCGs M87, NGC 4696, Hydra A and PKS0745-19, led the authors to argue that such magnetic fields should prevent the randomization of proton velocity dispersion and therefore could not explain the low polarization degree measured in their observations at $\sim 0.1\%$, similar to what we observe in the current ESPaDOnS observations.

Additionally, the known presence of dust within the foreground infalling galaxy known as the High-Velocity System (HVS - \citealt{yu2015high}, \citealt{rhea2025multiwavelength}), could lead to the presence of linear polarization through scattering of the filaments' light. Indeed, the first region observed with ESPaDOnS is located close to the dust lanes of the HVS visible in absorption in HST observations (\citealt{fabian_magnetic_2008}, \citealt{yu2015high} as well as Figure \ref{fig:hst_pointings}). It is therefore a possibility for dust scattering to lead to the observed linear polarization in the ESPaDOnS observations. Dust scattering would lead to a linear polarization degree of a percent or below for the continuum emission (i.e. \citealt{bianchi1995monte}, \citealt{peest2017polarization}), which is on the order of magnitude of our observed continuum polarization degree of $\sim 0.25 \%$. Moreover, clear absorption or drops associated with the position of our main emission lines indicate a lower level of polarization associated with the emission features suggesting that the polarization mechanism affects the continuum more effectively than the filaments emission, which is a feature of dust scattering in polarization. Likewise, \cite{marin2025spectropolarimetry} suggest and explore a similar polarization source through scattering for the optical emission from the central AGN of NGC 1275, observing similar features of an enhanced polarization degree for the continuum and decreased polarization at the position of the main emission lines. 

Moreover, the apparent stable and wavelength-independent angle of polarization at $5.3\pm25.7^\circ$ on average for both the continuum and emission lines, implying an horizontal preferred axis of electric field oscillation for the observed light and thus of the polarization vector, could be the geometrical result of orthogonal scattering of the light by dust grains along the line of sight. Similar properties of extremely low ($\sim 0^\circ$) measured polarization angles are also observed through spectropolarimetric studies of comets molecular emission and dust content (see \citealt{kwon2018high}). Unfortunately, the absence of clear detectable H$\beta$ emission, as visible in Appendix \ref{annex:hbeta_oiii}, prevents us from calculating a dust reddening correction factor, to estimate the level of dust within the aperture of Region 1. Finally, a key characteristic of polarization through dust scattering would be a wavelength dependant polarization degree, increasing at higher wavelengths (e.g. \citealt{serkowski1975wavelength}, \citealt{siebenmorgen2014dust}). However, due to the nature of the ESPaDOnS observations and spectral data, we are not able to determine such a property in the polarimetric data, as the overall spectral intensity is also significantly wavelength-dependant on large wavebands, thus preventing us from determining an accurate polarization degree across the spectrum. Nevertheless, scattering as a source of polarization within the filaments appears as the most likely origin given the current limited data available. 

Finally, the superposition of multiple emission and physical structures within the filamentary nebula, as explored in Section \ref{sec:superposition}, could affect the overall polarization profile of the resulting emission as observed with ESPaDOnS. Indeed, assuming a model where an outer filament's emission is associated with the narrow emission component and an AGN outflow or filaments churned by the currently forming radio bubbles could be linked to the broader emission components, this implies that the observed total emission is imprinted by significantly different ionizing and polarizing mechanisms which complicates our understanding of the observed polarization patterns. As described in Section \ref{sec:superposition}, a noticeable feature of the normalized q and u Stokes parameters is the velocity-dependant nature of the features and dips visible at the position of the multiple emission components of the H$\alpha$ emission line. Interestingly, when focusing first on the normalized u Stokes parameter displayed in the left panel of Figure \ref{fig:Stokes_parameters}, we can see that for the narrow blue component, an increase in u is visible at the peak of the emission but drops in its wings. Similarly, for the intermediate red component in the left panel of Figure \ref{fig:Stokes_parameters}, an increase in u appears noticeable at the peak of this component while drops are visible in its wings as well. This picture is less apparent for the broader but fainter green component, yet it overall appears to indicate that the multiple superimposed emission components each affect the polarization profile independently to a degree. Peculiarly, such a picture is not clearly visible in the normalized q Stokes parameter. Indeed, as seen in the right panel of Figure \ref{fig:Stokes_parameters}, we can observe that a smaller non-negligible dip is found in the blue-shifted wing of the H$\alpha$ emission line and appears centered at the location where the multiple emission components overlap. This, in turn, could indicate that the Stokes parameters and thus the polarization vector are affected by the filaments' emission and their combination. Therefore, as we interpreted in Section \ref{sec:superposition}, the spatial superposition of filaments, each with distinct emission and physical properties, could lead to noticeable velocity-dependant features in Stokes parameters, therefore affecting the overall observed polarization properties of the filaments within ESPaDOnS' aperture.

Nevertheless, the lack of polarimetric observations across more extended regions of the filaments surrounding NGC 1275 and other BCGs prevents us from drawing a decisive conclusion on a tentative source of polarization. However, the high-spectral-resolution ESPaDOnS observations presented here, offer a clear glimpse of the possible key role of polarimetric observations to further comprehend characteristics of the filaments and their environment. Moreover, recent magnetohydrodynamic simulations of hot and cold gas accreting onto the supermassive black holes of elliptical galaxies suggests that filaments of cool ionised gas are magnetized (e.g. \citealt{2024ApJ...973..141G}). Extended observations of circular polarization through the analysis of V Stokes parameter could possibly offer us the opportunity to detect and measure the presence of magnetic fields in these filamentary nebula around BCGs. 

\section{Conclusion} \label{sec:conclusion}

We analyzed archival high spectral resolution (R=68000) observations with ESPaDOnS of three pointings displaying clear emission lines, namely \sii$\lambda6716$, \sii$\lambda6731$, \nii$\lambda6584$, H$\alpha$(6563\AA), and \nii$\lambda6548$, associated with the filamentary nebula surrounding NGC 1275, the brightest cluster galaxy of the Perseus cluster. The polarimetric data associated with these observations also display clear features linked to the emission lines allowing us to derive for the first time a polarization degree and angle of polarization for the filaments surrounding NGC 1275.

\begin{itemize}
    \item We performed a complete MCMC fitting procedure of the three ESPaDOnS pointings allowing us, for the first time, to disentangle multiple emission features with precision and to determine their amplitude, velocity and velocity dispersion.

    \item ESPaDOnS' Region 1 thus shows three emission components with velocity dispersion of $\sim 25.1$ km/s, $\sim 89.6$ km/s and $\sim 207.6$ km/s respectively for the H$\alpha$ emission line, which appear similar to previous SITELLE high-spectral resolution (R=7000) observations of the filaments surrounding NGC 1275 where outer filaments appear extremely quiescent with an average velocity dispersion of $\sim 44$ km/s, while inner filaments display much broader emission with an average velocity dispersion of $\sim 140$ km/s (\citealt{vigneron2024high}).

    \item ESPaDOnS' Regions 2 \& 3 only appear to display a single emission component with relatively low SNR but slightly higher velocity dispersion ($\sim 118.3$ and $\sim 78.0$ km/s respectively) only allowing us to detect the \nii$\lambda6584$, H$\alpha$(6563\AA), and \nii$\lambda6548$ emission lines.

    \item The emission line intensity ratio analysis of Region 1 appears to indicate clear different sources of ionization for the three emission components found within its spectra. The narrow emission component demonstrates overall low \nii$\lambda6548$ by H$\alpha$ ratio, indicating a tentative ionization by collisional excitation, while also displaying a likely low electron density through its \sii$\lambda6716$ by \sii$\lambda6731$ ratio. The broadest component on the other hand shows a high \nii$\lambda6548$ by H$\alpha$ ratio indicative of a strong local ionizing source, possibly through AGN activity with a high electron density. Finally, the intermediate component shows a possibly mixed ionizing source and a likely low electron density as determined through the \nii$\lambda6548$ by H$\alpha$ and \sii$\lambda6716$ by \sii$\lambda6731$ ratios.
    
    \item On the other hand, the second and third pointing appear to display similar mixed to high energetic ionizing sources through their respective \nii$\lambda6548$ by H$\alpha$ ratios.

    \item We analyzed the associated polarimetric data for the first region of the ESPaDOnS observations allowing us to retrieve the Stokes Q and U parameters. We report the detection of a polarization degree of $\sim 0.15\%$ at the position of the H$\alpha$ emission line and a polarization degree of $\sim 0.26\%$ for the continuum. The angle of polarization is independent of wavelength and centered at $\sim 5.3\pm25.7^\circ$.

    \item We explored potential sources of polarization to explain these features through magnetic fields, dust scattering and the superposition of multiple physical structures possibly reinforcing the polarization of the overall emission, but the lack of more extensive polarimetric data prevents us from drawing a clear conclusion on the source of polarization for the filaments' emission, although scattering appears as the most likely candidate.
    
\end{itemize}

This article is based on observations obtained at the Canada-France-Hawai'i Telescope (CFHT) which is operated by the National Research Council of Canada, the Institut National des Sciences de l'Univers of the Centre National de la Recherche Scientifique of France, and the University of Hawai'i. CFHT is located on Maunakea on Hawai'i Island, a mountain of considerable cultural, natural, and ecological significance. Maunakea is a sacred site to Native Hawaiians, also known as Kānaka 'Ōiwi. We would like to thank the Canada-France-Hawai'i Telescope (CFHT) Operations and Software Groups for their contributions and diligence in maintaining observatory operations; the CFHT Astronomy Group for their observation coordination and data acquisition efforts; and the CFHT Finance \& Administration Group for their contributions to the management and administration of the observatory.

B.V. acknowledges financial support from the physics department of the Université de Montréal. J.H.L. acknowledges funding support from the Canada Research Chairs Program, as well as the Natural Sciences and Engineering Research Council of Canada (NSERC) through the Discovery Grant and Accelerator Supplement programs. J. B-B acknowledges support from project UNAM DGAPA-PAPIIT AG 101025, Mexico, as well as support from the PASPA 2025 grant. This work was supported by grant \href{https://doi.org/10.69777/377645}{https://doi.org/10.69777/377645} from the Fonds de recherche du Québec and by the Center for research in astrophysics of Québec (AstroQuébec).

\software{python (\citealt{van2009python}), astropy (\citealt{robitaille_astropy_2013}, \citealt{price-whelan_astropy_2018}, \citealt{2022ApJ...935..167A}), numpy (\citealt{harris_array_2020}), scipy (\citealt{virtanen_scipy_2020}), matplotlib (\citealt{hunter_matplotlib_2007}), sherpa (\citealt{2001SPIE.4477...76F}, \citealt{2007ASPC..376..543D}, \citealt{2024ApJS..274...43S}), \texttt{LUCI} (\citealt{rhea_luci_2021}), dynesty (\citealt{2020MNRAS.493.3132S}, \citealt{2004AIPC..735..395S}), emcee (\citealt{2013PASP..125..306F}).}

\clearpage
\appendix
\section{Other ESPaDOnS Pointings}
\label{annex:other_pointings}
We display below the other two pointings from the ESPaDOnS observations displaying no clear signal at the position of the H$\alpha$ and \nii\text{} emission lines and denoted as Regions 4 \& 5. The ESPaDOnS instrument uses offset pointing to perform its observation targeting and private communications with the instrument PIs informed us that the positioning of these two specific regions might have suffered from an incorrect pointing leading to the absence of emission detection, especially in the case of Region 5.

\begin{figure}[h!]
    \centering
    \includegraphics[width=0.85\linewidth]{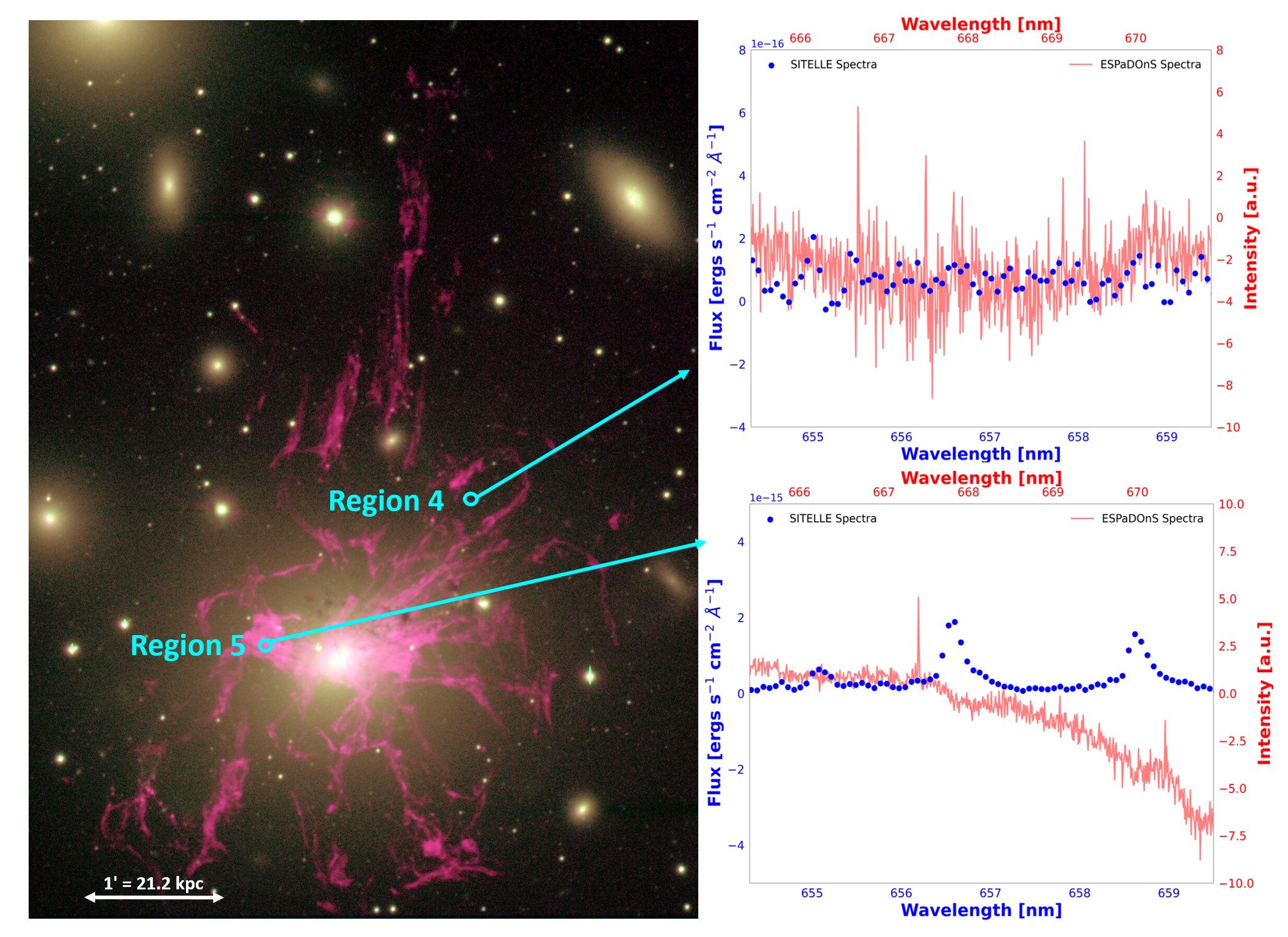}
    \caption{Location of the other two ESPaDOnS pointings displaying no detectable emission signal within the filamentary nebula surrounding NGC 1275 overplotted on WIYS observations (\citealt{conselice2001nature}). The associated ESPaDOnS and SITELLE spectra at high spectral resolution are shown in red and blue respectively with their corresponding location indicated with cyan circles denoted as Region 4, located at RA: 3:19:44.9 and DEC: +41:31:25.6, and Region 5, located at RA: 3:19:50.00 and DEC: +41:30:47.0. Since the SITELLE observations are redshift-corrected and the ESPaDOnS observations are not, the corresponding wavelengths axis associated with each dataset are also colour-coded (respectively in blue for SITELLE and red for ESPaDOnS).}
    \label{fig:other_pointings}
\end{figure}

\FloatBarrier
\section{H\texorpdfstring{$\beta$}\text{} and [O III] Emission Lines}
\label{annex:hbeta_oiii}

We display below the ESPaDOnS spectra of Region 1 at the position of the H$\beta$ and [O III] emission lines. As can be seen, we are unable to properly resolve any signal due to either absence of detection or the presence of instrumental artifacts. 

\begin{figure*}[hbt!]
\begin{subfigure}{0.49\linewidth}
  \centering
  \includegraphics[width=\linewidth]{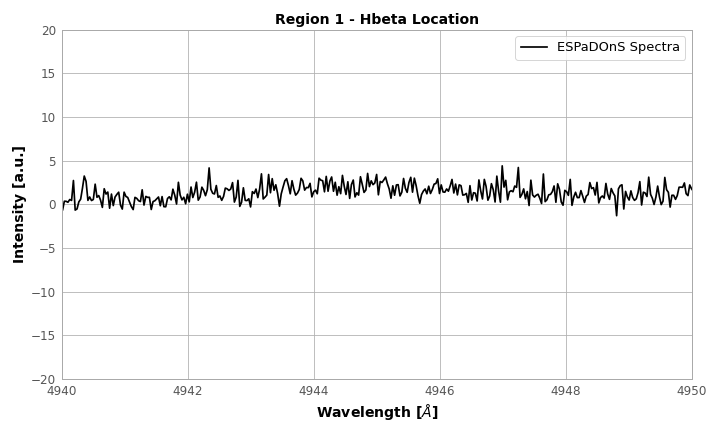}
  \label{hbeta}
\end{subfigure} 
\medskip
\begin{subfigure}{0.49\linewidth}
  \centering
  \includegraphics[width=\linewidth]{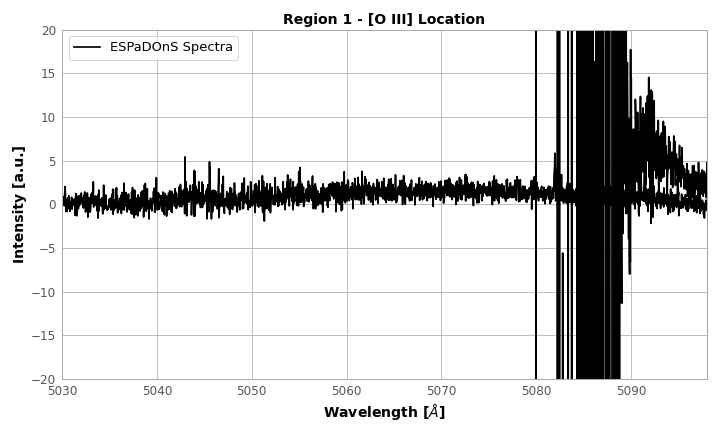}
  \label{oiii}
\end{subfigure}
\caption{Left: Emission spectra from the ESPaDOnS observations of Region 1 centered on the H$\beta$ emission line, where no clear detection is visible. Right: Emission spectra from the ESPaDOnS observations of Region 1 centered on the [O III] emission line doublet, where no clear detection is visible and an instrumental artifact severely impedes the emission detection.}
\label{fig:hbeta_oiii}
\end{figure*}

\FloatBarrier
\section{MCMC Priors}
\label{annex:mcmc_priors}
We display below in Table \ref{tab:mcmc_priors} the list of uniform priors used during the MCMC fitting procedure described in Section \ref{sec:fitting} using the \texttt{emcee} package.

\begin{deluxetable}{l l l r r r}
\tablecaption{MCMC Parameters Priors
\label{tab:mcmc_priors}}
\tablehead{
\colhead{Region} &
\colhead{Component} &
\colhead{Line} &
\colhead{FWHM priors} &
\colhead{Centroid priors} &
\colhead{Intensity priors} \\
&
&
&
\colhead{(km s$^{-1}$)} &
\colhead{(\AA)} &
\colhead{(a.u.)}
}

\startdata
&  & [\ion{S}{2}]$\lambda6716$ & $30 - 100$ & $6820 - 6840$ & $0.1-1.0$ \\
&  & [\ion{S}{2}]$\lambda6731$ & $30 - 100$ & $6835 - 6855$ & $0.1-3.0$ \\
& Narrow & [\ion{N}{2}]$\lambda6548$ & $30 - 100$ & $6655 - 6665$ & $0.1 - 4.0$ \\
&  & H$\alpha$ & $30 - 100$ & $6670 - 6680$ & $1.0-8.0$ \\
&  & [\ion{N}{2}]$\lambda6584$ & $30 - 100$ & $6690 - 6700$ & $0.1 - 4.0$ \\
\cline{2-6}
&  & [\ion{S}{2}]$\lambda6716$ & $100 - 300$ & $6820 - 6840$ & $1.0-7.0$ \\
&  & [\ion{S}{2}]$\lambda6731$ & $100 - 300$ & $6835 - 6855$ & $1.0-7.0$ \\
Region 1 & Intermediate & [\ion{N}{2}]$\lambda6548$ & $100 - 300$ & $6655 - 6665$ & $2.0-10.0$ \\
&  & H$\alpha$ & $100 - 300$ & $6670 - 6680$ & $10.0 - 50.0$ \\
&  & [\ion{N}{2}]$\lambda6584$ & $100 - 300$ & $6690 - 6700$ & $6.0-30.0$ \\
\cline{2-6}
&  & [\ion{S}{2}]$\lambda6716$ & $300 - 600$ & $6820 - 6840$ & $2.0-10.0$ \\
&  & [\ion{S}{2}]$\lambda6731$ & $300 - 600$ & $6835 - 6855$ & $2.0-10.0$ \\
& Broad & [\ion{N}{2}]$\lambda6548$ & $300 - 600$ & $6655 - 6665$ & $5.0-10.0$ \\
&  & H$\alpha$ & $300 - 600$ & $6670 - 6680$ & $15.0-30.0$ \\
&  & [\ion{N}{2}]$\lambda6584$ & $300 - 600$ & $6690 - 6700$ & $15.0-30.0$ \\
\hline
&  & [\ion{N}{2}]$\lambda6548$ & $100 - 500$ & $6655 - 6665$ & $1.0-6.0$ \\
Region 2 & Single & H$\alpha$ & $100 - 500$ & $6670 - 6680$ & $6.0-18.0$ \\
&  & [\ion{N}{2}]$\lambda6584$ & $100 - 500$ & $6690 - 6700$ & $3.0-18.0$ \\
\hline
&  & [\ion{N}{2}]$\lambda6548$ & $100 - 500$ & $6655 - 6665$ & $1.0-6.0$ \\
Region 3 & Single & H$\alpha$ & $100 - 500$ & $6670 - 6680$ & $6.0-18.0$ \\
&  & [\ion{N}{2}]$\lambda6584$ & $100 - 500$ & $6690 - 6700$ & $3.0-18.0$ \\
\enddata
\end{deluxetable}

\newpage


\bibliography{sample701}{}
\bibliographystyle{aasjournal}


\end{document}